\documentclass[conference]{IEEEtran}

\usepackage{subfigure}
\usepackage{graphicx}
\usepackage{amsmath}
\usepackage{bm}
\usepackage{url}
\usepackage{stmaryrd}
\usepackage{balance}
\usepackage{amssymb}
\usepackage{pgfplots}
\usepackage{pifont}
\usepackage{epsfig}
\usepackage{booktabs}
\usepackage{pgffor}
\usepackage{tikz}
\usepackage{multirow}
\usepackage{amsmath}
\usepackage{bbm}

\makeatletter
\newcommand\xleftrightarrow[2][]{%
  \ext@arrow 9999{\longleftrightarrowfill@}{#1}{#2}}
\newcommand\longleftrightarrowfill@{%
  \arrowfill@\leftarrow\relbar\rightarrow}
\makeatother

\newcommand{\mb}{\mathbf}

\newtheorem{definition}{\textsc{Definition}}

\newcommand{\problem}{fake news detection}
\newcommand{\our}{\textsc{FakeDetector}}
\newcommand{\rnn}{\textsc{Rnn}}
\newcommand{\deepwalk}{\textsc{DeepWalk}}
\newcommand{\modelline}{\textsc{Line}}

\newcommand{\svm}{\textsc{Svm}}
\newcommand{\propagation}{\textsc{Propagation}}
\newcommand{\unit}{\textsc{Gdu}}
\newcommand{\cnn}{Hybrid \textsc{Cnn}}
\newcommand{\liwc}{\textsc{Liwc}}
\newcommand{\trifn}{\textsc{TriFn}}

\usepackage{algorithm}
\usepackage{algorithmic}

\begin{document}

 



\title{{\our}: Effective Fake News Detection with Deep Diffusive Neural Network}



\author{Anonymous Submission}

\author{\IEEEauthorblockN{Jiawei Zhang$^1$, Bowen Dong$^2$, Philip S. Yu$^2$}
\IEEEauthorblockA{%
  $^1$IFM Lab, Department of Computer Science, Florida State University, FL, USA \\
  {$^2$BDSC Lab, Department of Computer Science, University of Illinois at Chicago, IL, USA}\\
  jzhang@ifmlab.org, \{bdong, psyu\}@uic.edu}}

\maketitle

\begin{abstract}

In recent years, due to the booming development of online social networks, fake news for various commercial and political purposes has been appearing in large numbers and widespread in the online world. With deceptive words, online social network users can get infected by these online fake news easily, which has brought about tremendous effects on the offline society already. An important goal in improving the trustworthiness of information in online social networks is to identify the fake news timely. This paper aims at investigating the principles, methodologies and algorithms for detecting fake news articles, creators and subjects from online social networks and evaluating the corresponding performance. This paper addresses the challenges introduced by the unknown characteristics of fake news and diverse connections among news articles, creators and subjects. This paper introduces a novel automatic fake news credibility inference model, namely {\our}. Based on a set of explicit and latent features extracted from the textual information, {\our} builds a deep diffusive network model to learn the representations of news articles, creators and subjects simultaneously. Extensive experiments have been done on a real-world fake news dataset to compare {\our} with several state-of-the-art models, and the experimental results have demonstrated the effectiveness of the proposed model.

\end{abstract}


\section{Introduction}\label{sec:introduction}

Fake news denotes a type of yellow press which intentionally presents misinformation or hoaxes spreading through both traditional print news media and recent online social media. Fake news has been existing for a long time, since the ``Great moon hoax'' published in 1835 \cite{great}. In recent years, due to the booming developments of online social networks, fake news for various commercial and political purposes has been appearing in large numbers and widespread in the online world. With deceptive words, online social network users can get infected by these online fake news easily, which has brought about tremendous effects on the offline society already. During the 2016 US president election, various kinds of fake news about the candidates widely spread in the online social networks, which may have a significant effect on the election results. According to a post-election statistical report \cite{AG17}, online social networks account for more than $41.8\%$ of the fake news data traffic in the election, which is much greater than the data traffic shares of both traditional TV/radio/print medium and online search engines respectively. An important goal in improving the trustworthiness of information in online social networks is to identify the fake news timely, which will be the main tasks studied in this paper. 


Fake news has significant differences compared with traditional suspicious information, like spams \cite{XWLY12_KDD, XWLY12_WWW, GHWLCZ10, ACF13}, in various aspects: (1) \textit{impact on society}: spams usually exist in personal emails or specific review websites and merely have a local impact on a small number of audiences, while the impact fake news in online social networks can be tremendous due to the massive user numbers globally, which is further boosted by the extensive information sharing and propagation among these users \cite{LHZY15, TTYC15, ZZWYX15}; (2) \textit{audiences' initiative}: instead of receiving spam emails passively, users in online social networks may seek for, receive and share news information actively with no sense about its correctness; and (3) \textit{identification difficulty}: via comparisons with abundant regular messages (in emails or review websites), spams are usually easier to be distinguished; meanwhile, identifying fake news with erroneous information is incredibly challenging, since it requires both tedious evidence-collecting and careful fact-checking due to the lack of other comparative news articles available.

These characteristics aforementioned of fake news pose new challenges on the detection task. Besides detecting fake news articles, identifying the fake news creators and subjects will actually be more important, which will help completely eradicate a large number of fake news from the origins in online social networks. Generally, for the news creators, besides the articles written by them, we are also able to retrieve his/her profile information from either the social network website or external knowledge libraries, e.g., Wikipedia or government-internal database, which will provide fundamental complementary information for his/her background check. Meanwhile, for the news subjects, we can also obtain its textual descriptions or other related information, which can be used as the foundations for news subject credibility inference. From a higher-level perspective, the tasks of fake news article, creator and subject detection are highly correlated, since the articles written from a trustworthy person should have a higher credibility, while the person who frequently posting unauthentic information will have a lower credibility on the other hand. Similar correlations can also be observed between news articles and news subjects. In the following part of this paper, without clear specifications, we will use the general \textit{fake news} term to denote the \textit{fake news articles}, \textit{creators} and \textit{subjects} by default.

\noindent \textbf{Problem Studied}: In this paper, we propose to study the {\problem} (including the articles, creators and subjects) problem in online social networks. Based on various types of heterogeneous information sources, including both textual contents/profile/descriptions and the authorship and article-subject relationships among them, we aim at identifying fake news from the online social networks simultaneously. We formulate the {\problem} problem as a credibility inference problem, where the real ones will have a higher credibility while unauthentic ones will have a lower one instead. 

The {\problem} problem is not easy to address due to the following reasons:
\begin{itemize}
\item \textit{Problem Formulation}: The {\problem} problem studied in this paper is a new research problem, and a formal definition and formulation of the problem is required and necessary before studying the problem.

\item \textit{Textual Information Usage}: For the news articles, creators and subjects, a set of their textual information about their contents, profiles and descriptions can be collected from the online social media. To capture signals revealing their credibility, an effective feature extraction and learning model will be needed. 

\item \textit{Heterogeneous Information Fusion}: In addition, as mentioned before, the credibility labels of news articles, creators and subjects have very strong correlations, which can be indicated by the authorship and article-subject relationships between them. An effective incorporation of such correlations in the framework learning will be helpful for more precise credibility inference results of fake news.
\end{itemize}

To resolve these challenges aforementioned, in this paper, we will introduce a new fake news detection framework, namely {\our}. In {\our}, the {\problem} problem is formulated as a credibility score inference problem, and {\our} aims at learning a prediction model to infer the credibility labels of news articles, creators and subjects simultaneously. {\our} deploys a new hybrid feature learning unit (HFLU) for learning the explicit and latent feature representations of news articles, creators and subjects respectively, and introduce a novel deep diffusive network model with the gated diffusive unit for the heterogeneous information fusion within the social networks.

The remaining paper is organized as follows. At first, we will introduce several important concepts and formulate the {\problem} problem in Section~\ref{sec:formulation}. Before introducing the proposed framework, we will provide a detailed analysis about fake news dataset in Section~\ref{sec:data_analysis}, which will provide useful signals for the framework building. The framework {\our} is introduced in Section~\ref{sec:method}, whose effectiveness will be evaluated in Section~\ref{sec:experiment}. Finally, we will talk about the related works in Section~\ref{sec:related_work} and conclude this paper in Section~\ref{sec:conclusion}.


\section{Terminology Definition and Problem Formulation} \label{sec:formulation}

In this section, we will introduce the definitions of several important concepts and provide the formulation of the studied problem.

\subsection{Terminology Definition}

In this paper, we will use the ``\textit{news article}'' concept in referring to the posts either written or shared by users in online social media, ``\textit{news subject}'' to represent the topics of these news articles, and ``\textit{news creator}'' concept to denote the set of users writing the news articles.

\begin{definition}
(News Article): News articles published in online social networks can be represented as set $\mathcal{N} = \{n_1, n_2, \cdots, n_m\}$. For each news article $n_i \in \mathcal{N}$, it can be represented as a tuple $n_i = (n_i^t, n_i^c)$, where the entries denote its \textit{textual content} and \textit{credibility label}, respectively.
\end{definition}

In the above definition, the news article credibility label is from set $\mathcal{Y}$, i.e., $n_i^c \in \mathcal{Y}$ for $n_i \in \mathcal{N}$. For the PolitiFact dataset to be introduced later, its label set $\mathcal{Y} = $\{True, Mostly True, Half True, Mostly False, False, Pants on Fire!\} contains $6$ different class labels, whose credibility ranks from high to low. In addition, the news articles in online social networks are also usually about some topics, which are also called the \textit{news subjects} in this paper. News subjects usually denote the central ideas of news articles, and they are also the main objectives of writing these news articles.

\begin{definition}
(News Subject): Formally, we can represent the set of news subjects involved in the social network as $\mathcal{S} = \{s_1, s_2, \cdots, s_k\}$. For each subject $s_i \in \mathcal{S}$, it can be represented as a tuple $s_i = (s_i^t, s_i^c)$ containing its \textit{textual description} and \textit{credibility label}, respectively.
\end{definition}

\begin{definition}
(News Creator): We can represent the set of news creators in the social network as $\mathcal{U} = \{u_1, u_2, \cdots, u_n\}$. To be consistent with the definition of \textit{news articles}, we can also represent news creator $u_i \in \mathcal{U}$ as a tuple $u_i=(u_i^p, u_i^s)$, where the entries denote the \textit{profile information} and \textit{credibility label} of the creator, respectively. 
\end{definition}

For the news article creator $u_i \in \mathcal{U}$, his/her profile information can be represented as a sequence of words describing his/her basic background. For some of the creators, we can also have his/her title representing either their jobs, political party membership, their geographical residential locations or companies they work at, e.g., ``political analyst'', ``Democrat''/``Republican'', ``New York''/``Illinois'' or ``CNN''/``Fox''. Similarly, the credibility labels of the creator $u_i$ can also be assigned with a class label from set $\mathcal{Y}$.

\begin{definition}
(News Augmented Heterogeneous Social Networks): The online social network together with the news articles published in it can be represented as a \textit{news augmented heterogeneous social network} (News-HSN) $G = (\mathcal{V}, \mathcal{E})$, where the node set $\mathcal{V} = \mathcal{U} \cup \mathcal{N} \cup  \mathcal{S}$ covers the sets of news articles, creators and subjects, and the edge set $\mathcal{E} = \mathcal{E}_{u,n} \cup \mathcal{E}_{n,s}$ involves the authorship links between news articles and news creators, and the topic indication links between news articles and news subjects.
\end{definition}

\subsection{Problem Formulation}

Based on the definitions of terminologies introduced above, the {\problem} problem studied in this paper can be formally defined as follows.

\noindent \textbf{Problem Formulation}: Given a News-HSN $G = (\mathcal{V}, \mathcal{E})$, the {\problem} problem aims at learning an inference function $f: \mathcal{U} \cup \mathcal{N} \cup \mathcal{S} \to \mathcal{Y}$ to predict the \textit{credibility labels} of news articles in set $\mathcal{N}$, news creators in set $\mathcal{U}$ and news subjects in set $\mathcal{S}$. In learning function $f$, various kinds of heterogeneous information in network $G$ should be effectively incorporated, including both the textual content/profile/description information as well as the connections among them.


\section{Dataset Analysis}\label{sec:data_analysis}

The dataset used in this paper includes both the tweets posted by PolitiFact at its official Twitter account\footnote{https://twitter.com/PolitiFact}, as well as the fact-check articles written regarding these statements in the PolitiFact website\footnote{http://www.politifact.com}. In this section, we will first provide the basic statistical information about the crawled dataset, after which we will carry out a detailed analysis about the information regarding the news articles, creators and subjects, respectively.

\subsection{Dataset Statistical Information}

PolitiFact website is operated by the Tampa Bay Times, where the reporters and editors can make fact-check regarding the statements (i.e., the news articles in this paper) made by the Congress members, White House, lobbyists and other political groups (namely the ``creators'' in this paper). PolitiFact collects the political statements from the speech, news article report, online social media, etc., and will publish both the original statements, evaluation results, and the complete fact-check report at both PolitiFact website and via its official Twitter account. The statement evaluation results will clearly indicate the credibility rating, ranging from ``True'' for completely accurate statements to ``Pants on Fire!'' for totally false claims. In addition, PolitiFact also categorizes the statements into different groups regarding the subjects, which denote the topics that those statements are about. Based on the credibility of statements made by the creators, PolitiFact also provides the credibility evaluation for these creators and subjects as well. The crawled PolitiFact dataset can be organized as a network, involving articles, creators and subjects as the nodes, as well as the authorship link (between articles and creators) and subject indication link (between articles and subjects) as the connections. More detailed statistical information is provided as follows and in Table~\ref{tab:datastat}.

\begin{table}[t]
\caption{Properties of the Heterogeneous Networks}
\label{tab:datastat}
\centering
\begin{tabular}{clr}
\toprule
&property &\textbf{PolitiFact Network}   \\
\midrule 
\multirow{3}{*}{\# node}
&articles	&14,055 \\
&creators	&3,634 \\
&subjects	&152 \\
\midrule 
\multirow{3}{*}{\# link}
&creator-article	&14,055 \\
&article-subject	&48,756 \\
\bottomrule
\end{tabular}
\end{table}

From the Twitter account of PolitiFact, we crawled 14,055 tweets with fact check, which will compose the news article set in our studies. These news articles are created by $3,634$ creators, and each creator has created $3.86$ articles on average. These articles belong to $152$ subjects respectively, and each article may belong to multiple subjects simultaneously. In the crawled dataset, the number of article-subject link is $48,756$. On average, each news article has about $3.5$ associated subjects. Each news article also has a ``Truth-O-Meter'' rating score indicating its credibility, which takes values from \{True, Mostly True, Half True, Half False, Mostly False, Pants on Fire!\}. In addition, from the PolitiFact website, we also crawled the fact check reports on these news articles to illustrate the reasons why these news articles are real or fake, information from which is not used in this paper.

\subsection{Dataset Detailed Analysis}

Here, we will introduce a detailed analysis of the crawled PolitiFact network dataset, which can provide necessary motivations and foundations for our proposed model to be introduced in the next section. The data analysis in this section includes $4$ main parts: \textit{article credibility analysis with textual content}, \textit{creator credibility analysis}, \textit{creator-article publishing historical records}, as well as \textit{subject credibility analysis}, and the results are illustrated in Figure~\ref{fig:analysis} respectively.

\subsubsection{Article Credibility Analysis with Textual Content} 
\begin{figure*}[t]
\centering
\subfigure[Frequent Words used in True Articles.]{\label{fig:true_words}
    \begin{minipage}[l]{0.9\columnwidth}
      \centering
      \includegraphics[width=1.0\textwidth]{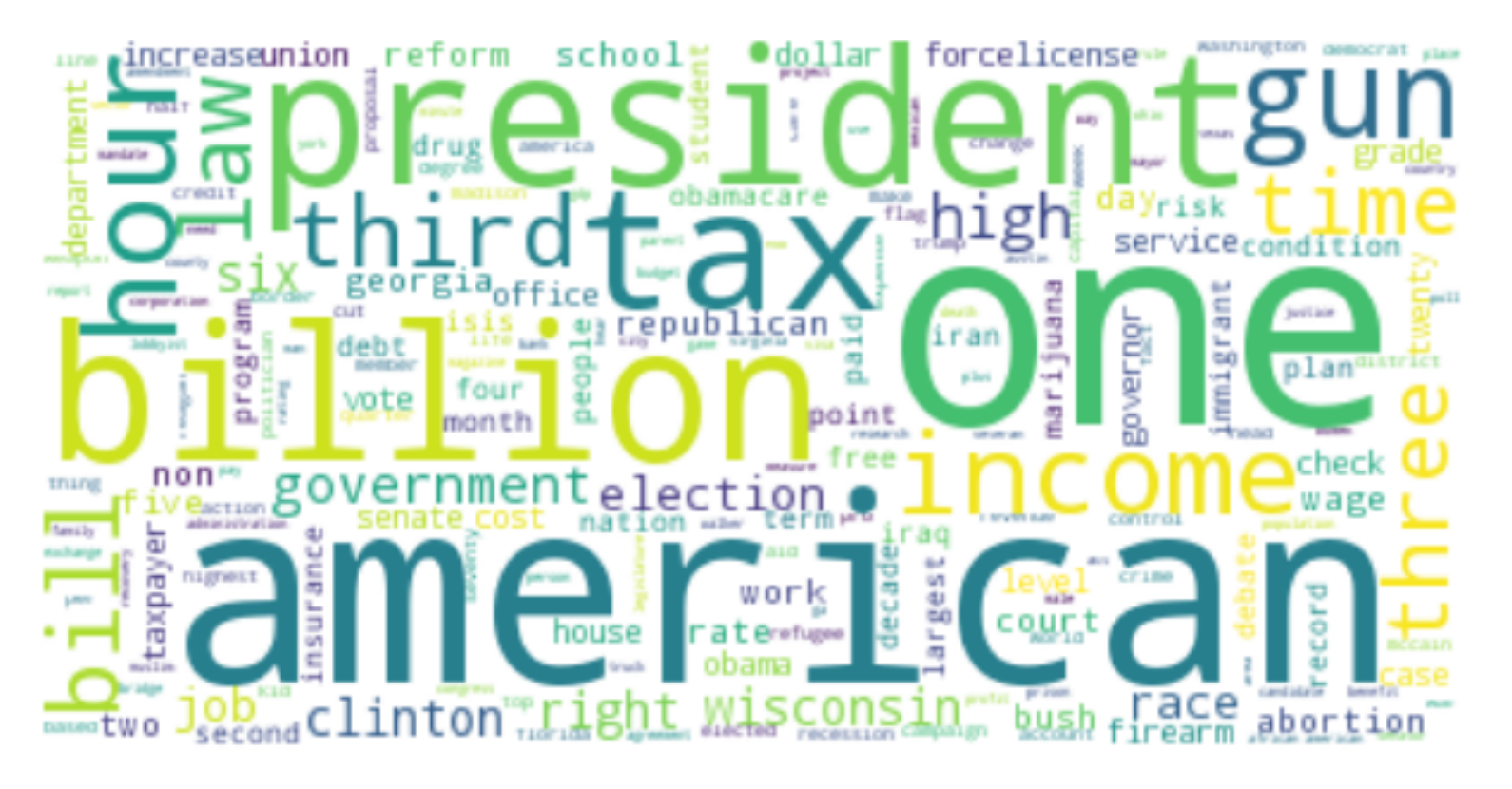}
    \end{minipage}
}
\subfigure[Frequent Words used in Fake Articles.]{\label{fig:false_words}
    \begin{minipage}[l]{0.9\columnwidth}
      \centering
      \includegraphics[width=1.0\textwidth]{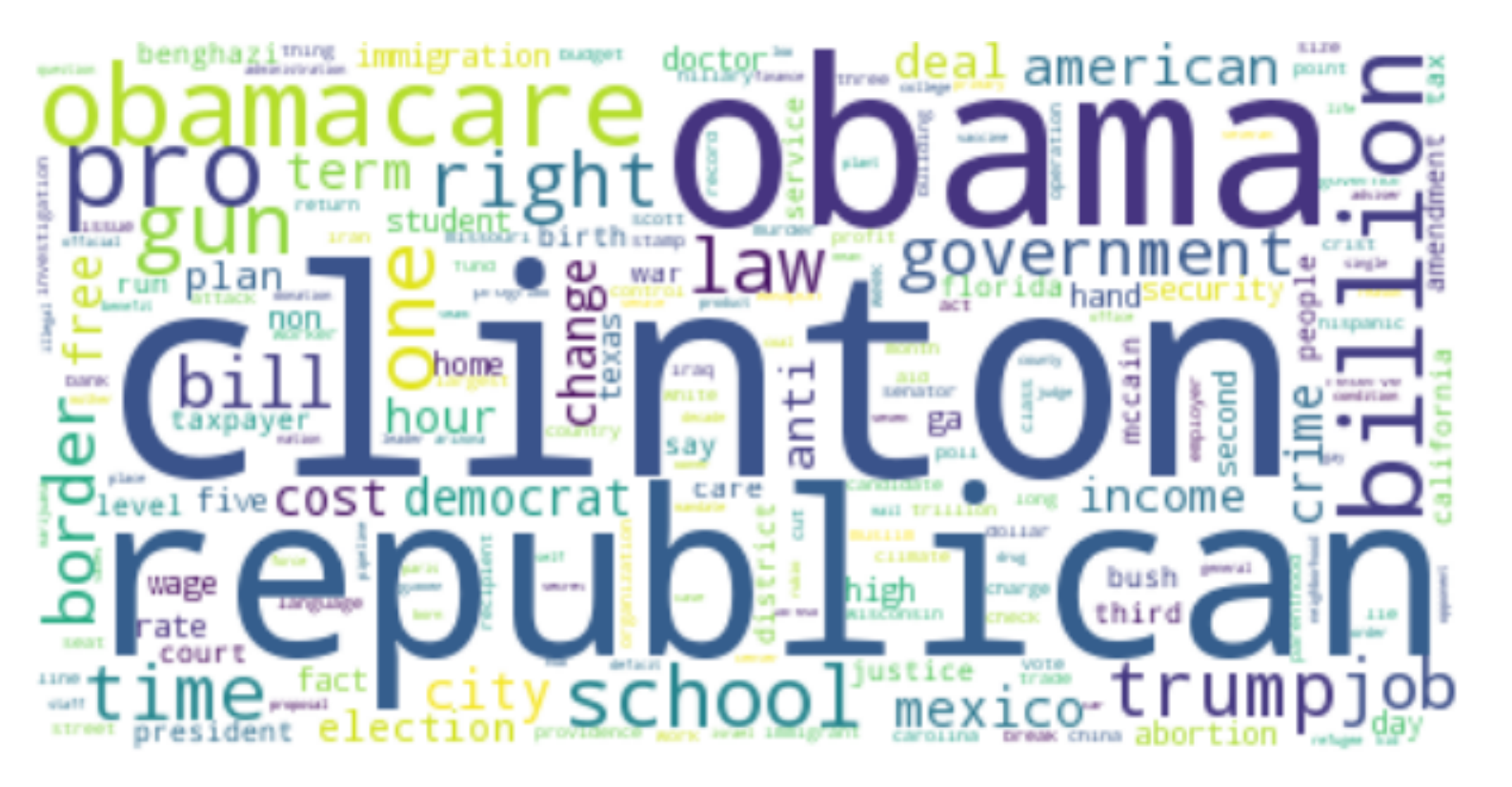}
    \end{minipage}
}
\subfigure[News Articles from the Republican.]{\label{fig:republican}
    \begin{minipage}[l]{0.9\columnwidth}
      \centering
      \includegraphics[width=1.0\textwidth]{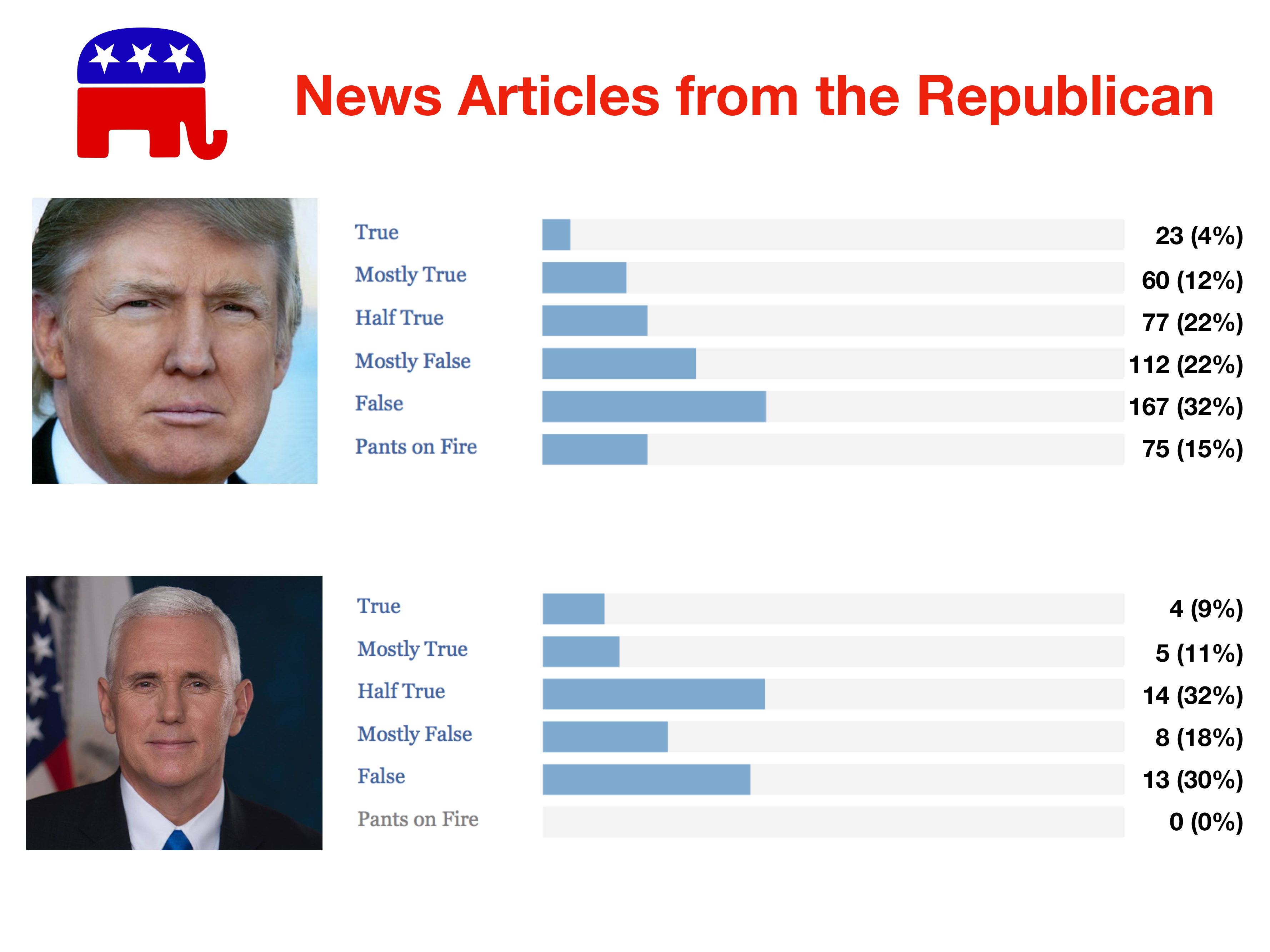}
    \end{minipage}
}
\subfigure[News Articles from the Democratic.]{\label{fig:democratic}
    \begin{minipage}[l]{0.9\columnwidth}
      \centering
      \includegraphics[width=1.0\textwidth]{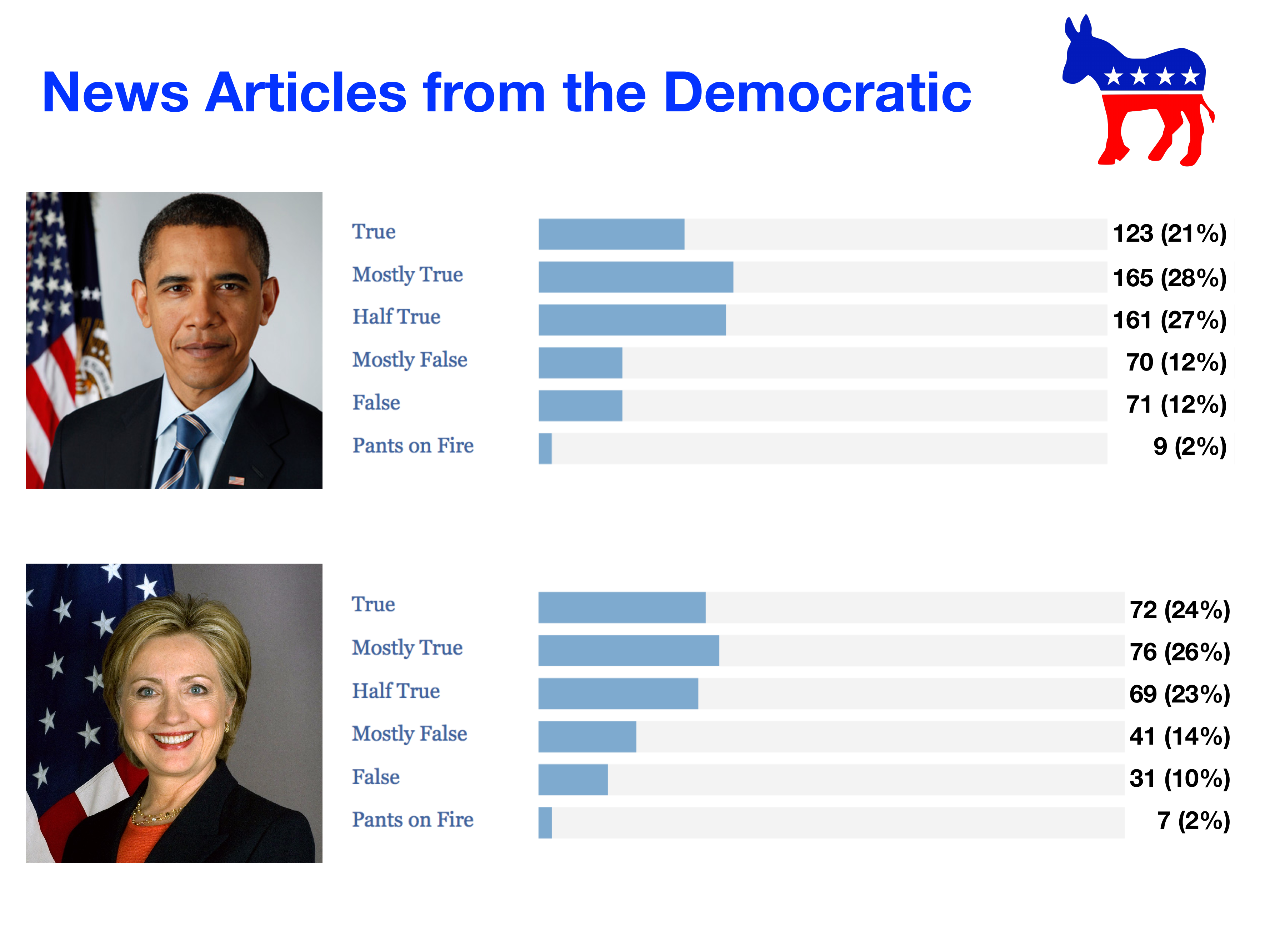}
    \end{minipage}
}
\subfigure[Power Law Distribution]{\label{fig:power_law}
    \begin{minipage}[l]{0.75\columnwidth}
      \centering
      \includegraphics[width=1.0\textwidth]{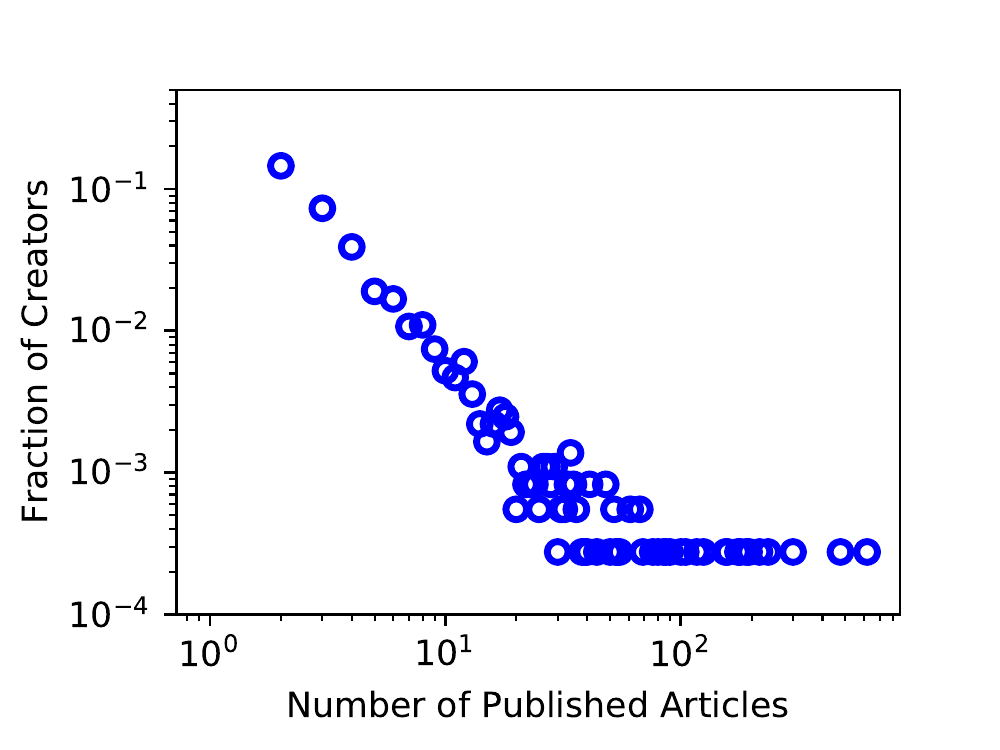}
    \end{minipage}
}
\subfigure[Subject Credibility Distribution.]{\label{fig:subject}
    \begin{minipage}[l]{0.75\columnwidth}
      \centering
      \includegraphics[width=1.0\textwidth]{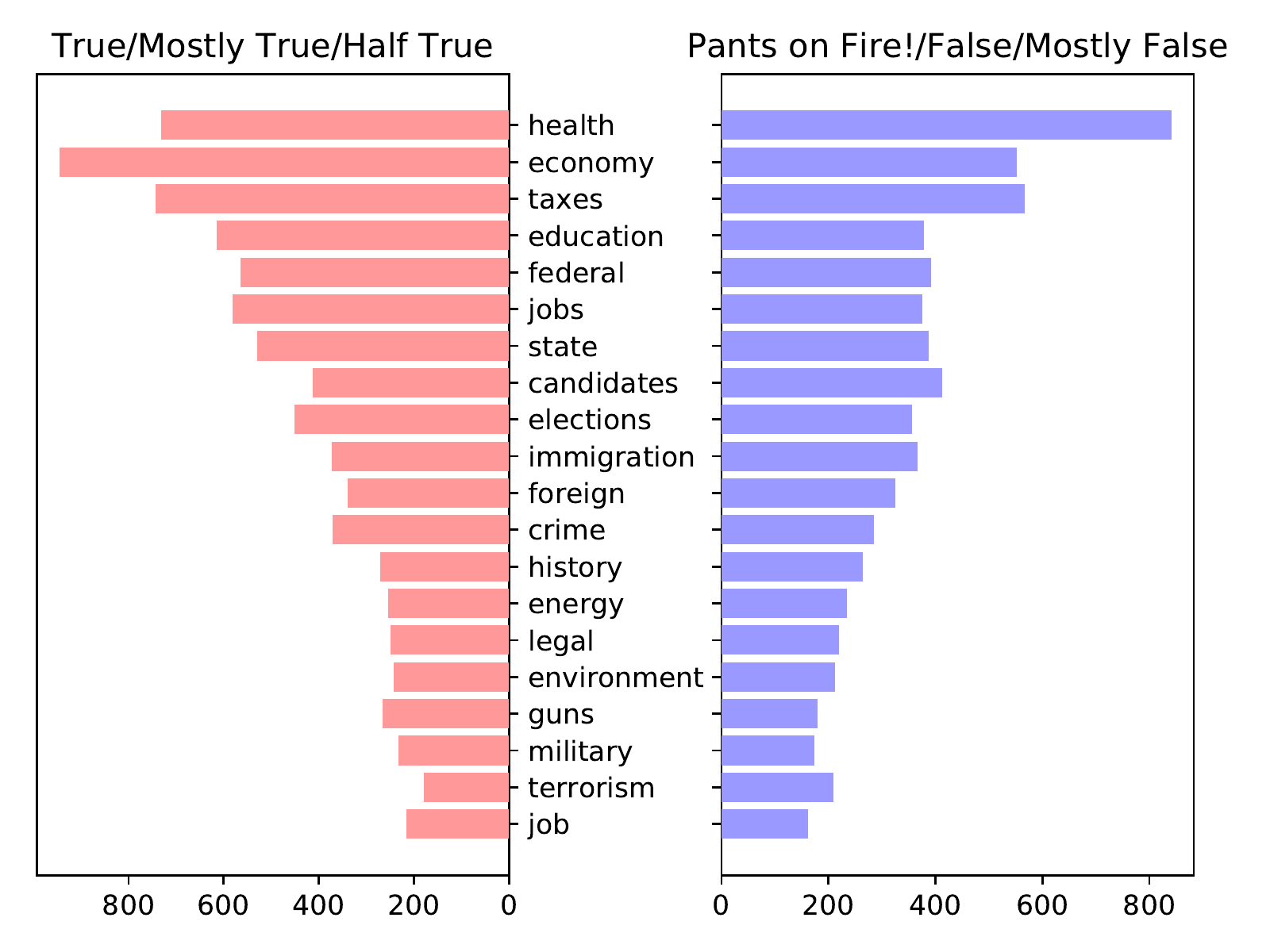}
    \end{minipage}
}
\caption{PolitiFact Dataset Statistical Analysis. (Plots (a)-(b): frequent words used in true/fake news articles; Plots (c)-(d): credibility of news articles from the Republican and the Democratic; Plots (e)-(f): some basic statistical information of the dataset.)}\label{fig:analysis}
\end{figure*}

In Figures~\ref{fig:true_words}-\ref{fig:false_words}, we illustrate the frequent word cloud of the true and false news articles, where the stop words have been removed already. Here, the true article set covers the news articles which are rated ``True'', ``Mostly True'' or ``Half True''; meanwhile, the false article set covers the news articles which are rated ``Pants on Fire!'', ``False'' or ``Mostly False''. According to the plots, from Figure~\ref{fig:true_words}, we can find some unique words in True-labeled articles which don't appear often in Figure~\ref{fig:false_words}, like ``President'', ``income'', ``tax'' and ``american'', ect.; meanwhile, from Figure~\ref{fig:false_words}, we can observe some unique words appearing often in the false articles, which include ``Obama'', ``republican'' ``Clinton'', ``obamacare'' and ``gun'', but don't appear frequently in the True-labeled articles. These textual words can provide important signals for distinguishing the true articles from the false ones.

\subsubsection{Creator Credibility Analysis} Meanwhile, in Figures~\ref{fig:republican}-\ref{fig:democratic}, we show $4$ case studies regarding the creators' credibility based on their published articles. We divide the case studies into two groups: republican vs democratic, where the representatives are ``Donald Trump'', ``Mike Pence'', ``Barack Obama'' and ``Hillary Clinton'', respectively. According to the plots, for the articles in the crawled dataset, most of the articles from ``Donald Trump'' in the dataset are evaluated to be false, which account for about $69\%$ of all his statements. For ``Mike Pence'', the ratio of true articles vs false articles is $52\%:48\%$ instead. Meanwhile, for most of the articles in the dataset from ``Barack Obama'' and ``Hillary Clinton'' are evaluated to be true with fact check, which takes more than $76\%$ and $73\%$ of their total articles respectively. 

\subsubsection{Creator-Article Publishing Historical Records} In Figure~\ref{fig:power_law}, we show the scatter plot about the distribution of the number of news articles regarding the fraction of creators who have published these numbers of articles in the dataset. According to the plot, the creator-article publishing records follow the power law distribution. For a large proportion of the creators, they have merely published less than $10$ articles, and a very small number of creators have ever published more than $100$ articles. Among all the creators, Barack Obama has the most articles, whose number is about $599$.

\subsubsection{Subject Credibility Analysis} Finally, in Figure~\ref{fig:subject}, we provide the statistics about the top 20 subjects with the largest number of articles, where the red bar denotes the true articles belonging to these subjects and the blue bar corresponds to the false news articles instead. According to the plot, among all the $152$ subjects, subject ``health'' covers the largest number of articles, whose number is about $1,572$. Among these articles, $731$ ($46.5\%$) of them are the true articles and $841$ ($53.5\%$) of them are false, and articles in this subject are heavily inclined towards the false group. The second largest subject is ``economy'' with $1,498$ articles in total, among which $946$ ($63.2\%$) are true and $552$ ($36.8\%$) are false. Different from ``health'', the articles belonging to the ``economy'' subject are biased to be true instead. Among all the top 20 subjects, most of them are about the economic and livelihood issues, which are also the main topics that presidential candidates will debate about during the election.

We need to add a remark here, the above observations are based on the crawled PolitiFact dataset only, and it only stands for the political position of the PolitiFact website service provider. Based on these observations, we will build a unified credibility inference model to identify the fake news articles, creators and subjects simultaneously from the network with a deep diffusive network model in the next section.

\section{Proposed Methods}\label{sec:method}

In this section, we will provide the detailed information about the {\our} framework in this section. Framework {\our} covers two main components: \textit{representation feature learning}, and \textit{credibility label inference}, which together will compose the deep diffusive network model {\our}.

\subsection{Notations}

Before talking about the framework architecture, we will first introduce the notations used in this paper. In the sequel of this paper, we will use the lower case letters (e.g., $x$) to represent scalars, lower case bold letters (e.g., $\mb{x}$) to denote column vectors, bold-face upper case letters (e.g., $\mb{X}$) to denote matrices, and upper case calligraphic letters (e.g., $\mathcal{X}$) to denote sets. Given a matrix $\mb{X}$, we denote $\mb{X}(i,:)$ and $\mb{X}(:,j)$ as the $i_{th}$ row and $j_{th}$ column of matrix $\mb{X}$ respectively. The ($i_{th}$, $j_{th}$) entry of matrix $\mb{X}$ can be denoted as either $X(i,j)$ or $X_{i,j}$, which will be used interchangeably in this paper. We use $\mb{X}^\top$ and $\mb{x}^\top$ to represent the transpose of matrix $\mb{X}$ and vector $\mb{x}$. For vector $\mb{x}$, we represent its $L_p$-norm as $\left\| \mb{x} \right\|_p = (\sum_i |x_i|^p)^{\frac{1}{p}}$. The $L_p$-norm of matrix $\mb{X}$ can be represented as $\left\| \mb{X} \right\|_p = (\sum_{i,j} |X_{i,j}|^p)^{\frac{1}{p}}$. The element-wise product of vectors $\mb{x}$ and $\mb{y}$ of the same dimension is represented as $\mb{x} \odot \mb{y}$, while the entry-wise plus and minus operations between vectors $\mb{x}$ and $\mb{y}$ are represented as $\mb{x} \oplus \mb{y}$ and $\mb{x} \ominus \mb{y}$, respectively.

\subsection{Representation Feature Learning}

As illustrated in the analysis provided in the previous section, in the news augmented heterogeneous social network, both the textual contents and the diverse relationships among news articles, creators and subjects can provide important information for inferring the credibility labels of fake news. In this part, we will focus on feature learning from the textual content information based on the \textit{hybrid feature extraction unit} as shown in Figure~\ref{fig:feature_extraction}, while the relationships will be used for building the deep diffusive model in the following subsection.

\subsubsection{Explicit Feature Extraction}

The textual information of fake news can reveal important signals for their credibility inference. Besides some shared words used in both true and false articles (or creators/subjects), a set of frequently used words can also be extracted from the article contents, creator profiles and subject descriptions of each category respectively. Let $\mathcal{W}$ denotes the complete vocabulary set used in the PolitiFact dataset, and from $\mathcal{W}$ a set of unique words can also be extracted from articles, creator profile and subject textual information, which can be denoted as sets $\mathcal{W}_{n} \subset \mathcal{W}$, $\mathcal{W}_{u} \subset \mathcal{W}$ and $\mathcal{W}_{s} \subset \mathcal{W}$ respectively (of size $d$). 

\begin{figure}[t]
\centering
    \begin{minipage}[l]{0.9\columnwidth}
      \centering
      \includegraphics[width=1.0\textwidth]{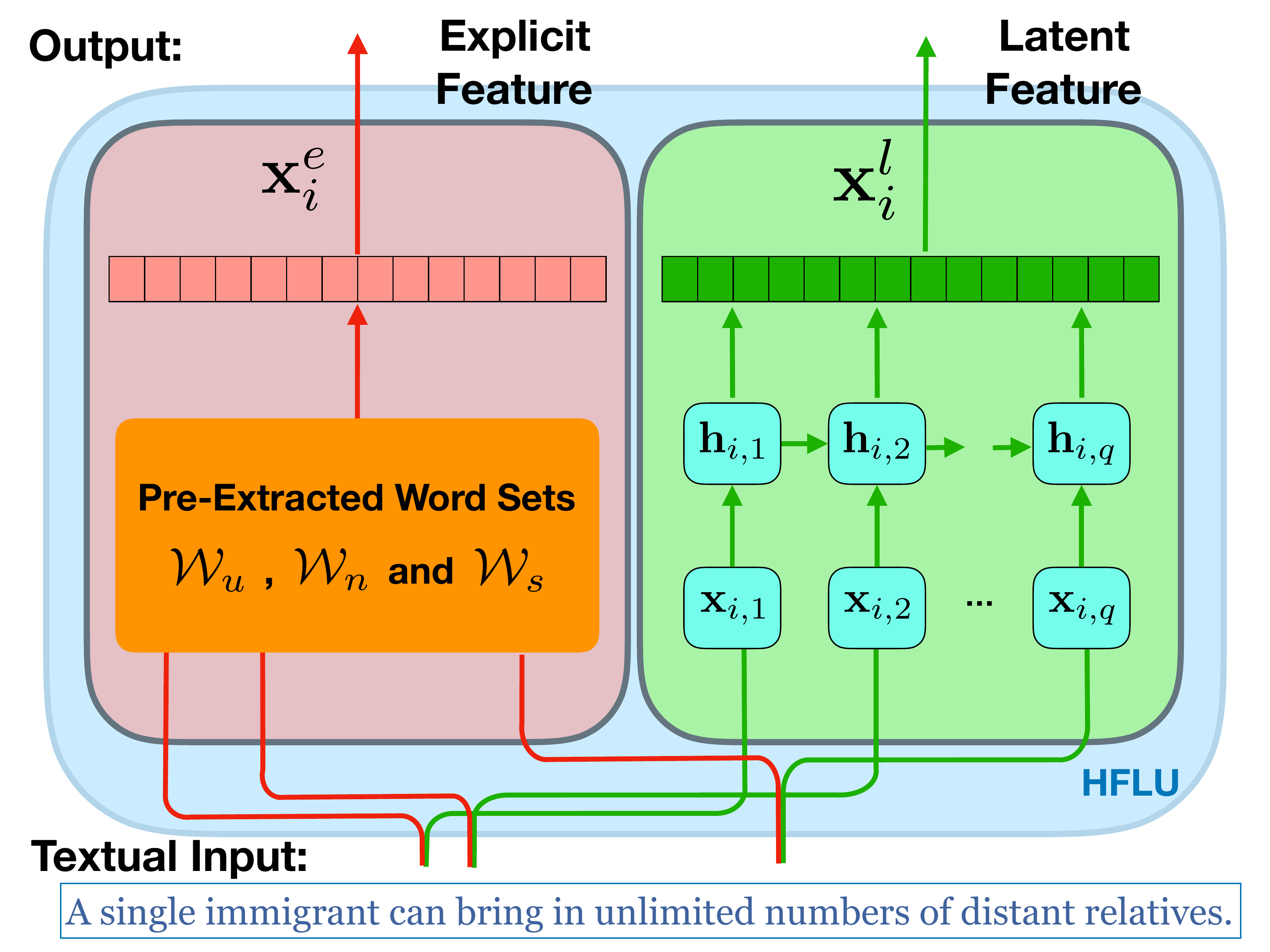}
    \end{minipage}
\caption{Hybrid Feature Learning Unit (HFLU).}\label{fig:feature_extraction}
\end{figure}

These extracted words have shown their stronger correlations with their fake/true labels. As shown in the left component of Figure~\ref{fig:feature_extraction}, based on the pre-extracted word sets $\mb{W}_{n}$, given a news article $n_i \in \mathcal{N}$, we can represent the extracted explicit feature vector for $n_i$ as vector $\mb{x}_{n,i}^e \in \mathbb{R}^d$, where entry ${x}_{n,i}^e(k)$ denotes the number of appearance times of word $w_k \in \mb{W}_{n}$ in news article $n_i$. In a similar way, based on the extracted word set $\mathcal{W}_{u}$ (and $\mathcal{W}_{s}$), we can also represent the extracted explicit feature vectors for creator $u_j$ as $\mb{x}_{u,j}^e \in \mathbb{R}^d$ (and for subject $s_l$ as $\mb{x}_{s,l}^e \in \mathbb{R}^d$). 

\subsubsection{Latent Feature Extraction}

Besides those explicitly visible words about the news article content, creator profile and subject description, there also exist some hidden signals about articles, creators and subjects, e.g., \textit{news article content information inconsistency} and \textit{profile/description latent patterns}, which can be effectively detected from the latent features as introduced in \cite{K14}. Based on such an intuition, in this paper, we propose to further extract a set of \textit{latent features} for news articles, creators and subjects based on the deep recurrent neural network model. 

Formally, given a news article $n_i \in \mathcal{N}$, based on its original textual contents, we can represents its content as a sequence of words represented as vectors $({\mb{x}}_{i,1}, {\mb{x}}_{i,2}, \cdots, {\mb{x}}_{i,q})$, where $q$ denotes the maximum length of articles (and for those with less than $q$ words, zero-padding will be adopted). Each feature vector ${\mb{x}}_{i,k}$ corresponds to one word in the article. Based on the vocabulary set $\mathcal{W}$, it can be represented in different ways, e.g., the one-hot code representation or the binary code vector of a unique index assigned for the word. The latter representation will save the computational space cost greatly.

As shown in the right component of Figure~\ref{fig:feature_extraction}, the latent feature extraction is based on RNN model (with the basic neuron cells), which has 3 layers (1 input layer, 1 hidden layer, and 1 fusion layer). Based on the input vectors $({\mb{x}}_{i,1}, {\mb{x}}_{i,2}, \cdots, {\mb{x}}_{i,q})$ of the textual input string, we can represent the feature vectors at the hidden layer and the output layer as follows respectively:
\begin{align*}
\begin{cases}
\hspace{-5pt} &\mbox{\# Fusion Layer:  } \mb{x}_{n,i}^l = \sigma(\sum_{t = 1}^q \mb{W}_{i} \mb{h}_{i,t}),\\
\hspace{-5pt} &\mbox{\# Hidden Layer:  } \mb{h}_{i,t} = GRU( \mb{h}_{i,t-1}, \mb{x}_{i,t}; \mb{W} )
\end{cases}
\end{align*}
where GRU (Gated Recurrent Unit) \cite{DBLP:journals/corr/ChungGCB14} is used as the unit model in the hidden layer and the $\mb{W}$ matrices denote the variables of the model to be learned.

Based on a component with a similar architecture, we can extract the latent feature vector for news creator $u_j \in \mathcal{U}$ (and subject $s_l \in \mathcal{S}$) as well, which can be denoted as vector $\mb{x}_{u,j}^l $ (and $\mb{x}_{s,l}^l $). By appending the explicit and latent feature vectors together, we can formally represent the extracted feature representations of news articles, creators and subjects as $\mb{x}_{n,i} = \left[(\mb{x}_{n,i}^e)^\top, (\mb{x}_{n,i}^l)^\top \right]^\top$, $\mb{x}_{u,j} = \left[(\mb{x}_{u,j}^e)^\top, (\mb{x}_{u,j}^l)^\top \right]^\top$ and $\mb{x}_{s,l} = \left[(\mb{x}_{s,l}^e)^\top, (\mb{x}_{s,l}^l)^\top \right]^\top$ respectively, which will be fed as the inputs for the deep diffusive unit model to be introduced in the next subsection.

\begin{figure}[t]
\centering
    \begin{minipage}[l]{0.9\columnwidth}
      \centering
      \includegraphics[width=1.0\textwidth]{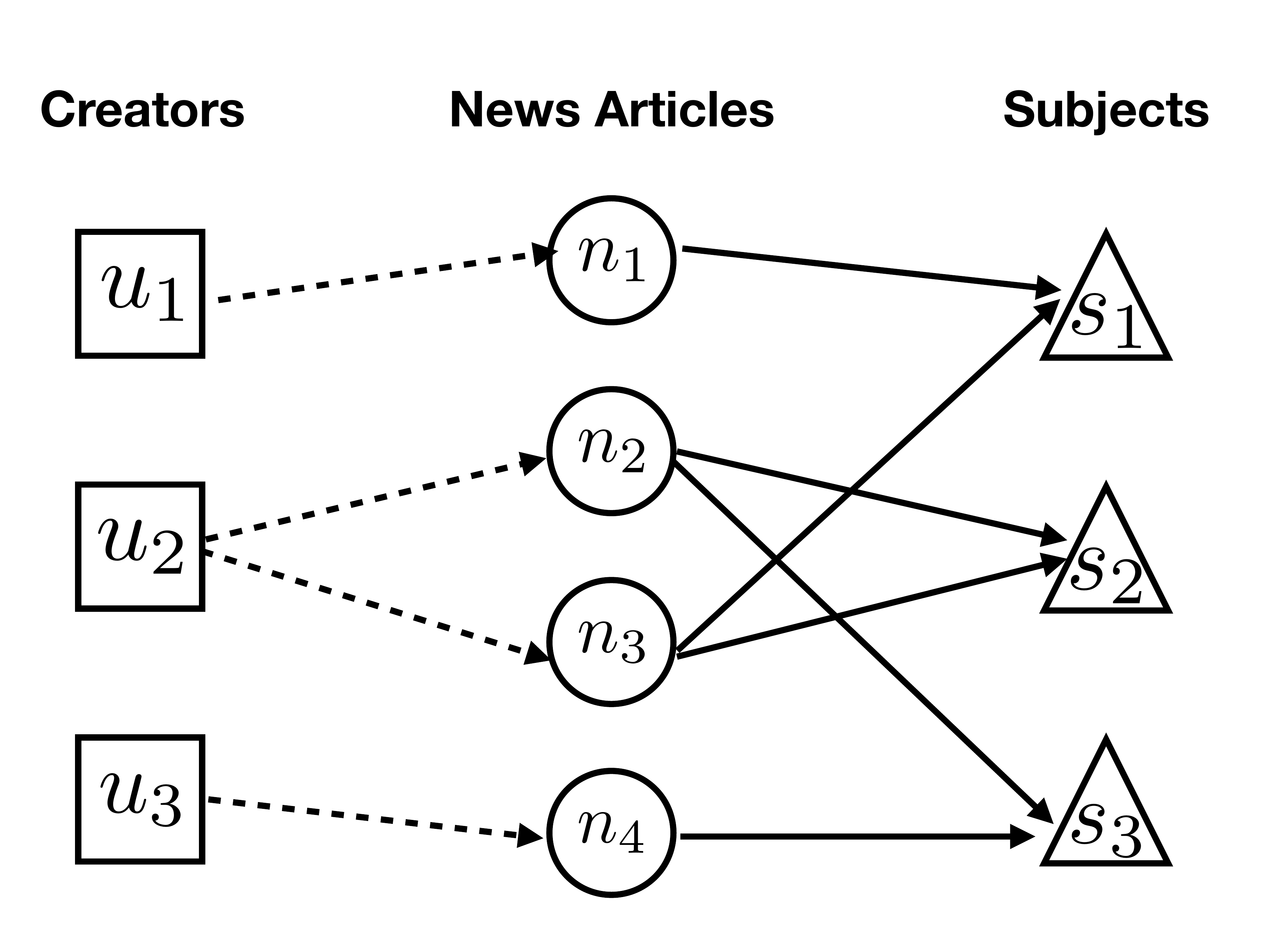}
    \end{minipage}
\caption{Relationships of Articles, Creators and Subjects.}\label{fig:relationship}
\end{figure}

\subsection{Deep Diffusive Unit Model}

Actually, the credibility of news articles are highly correlated with their subjects and creators. The relationships among news articles, creators and subjects are illustrated with an example in Figure~\ref{fig:relationship}. For each creator, they can write multiple news articles, and each news article has only one creator. Each news article can belong to multiple subjects, and each subject can also have multiple news articles taking it as the main topics. To model the correlation among news articles, creators and subjects, we will introduce the deep diffusive network model as follow.

\begin{figure}[t]
\centering
    \begin{minipage}[l]{0.9\columnwidth}
      \centering
      \includegraphics[width=1.0\textwidth]{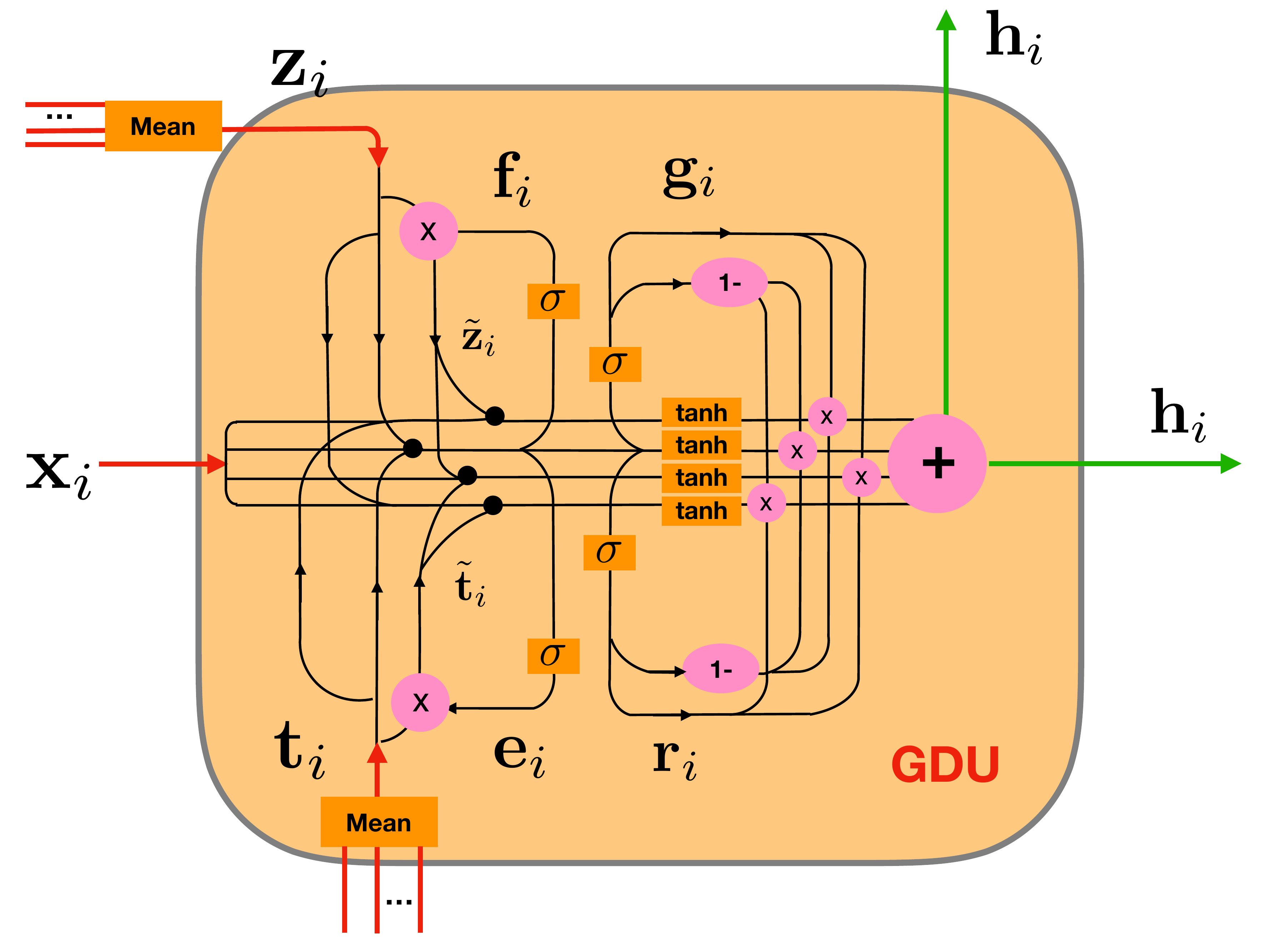}
    \end{minipage}
\caption{Gated Diffusive Unit (GDU).}\label{fig:unit}
\end{figure}

The overall architecture of {\our} corresponding to the case study shown in Figure~\ref{fig:relationship} is provided in Figure~\ref{fig:architecture}. Besides the HFLU feature learning unit model, {\our} also uses a \textit{gated diffusive unit} (GDU) model for effective relationship modeling among news articles, creators and subjects, whose structure is illustrated in Figure~\ref{fig:unit}. Formally, the GDU model accepts multiple inputs from different sources simultaneously, i.e., $\mb{x}_i$, $\mb{z}_i$ and $\mb{t}_i$, and outputs its learned hidden state $\mb{h}_i$ to the output layer and other unit models in the diffusive network architecture.

Here, let's take news articles as an example. Formally, among all the inputs of the GDU model, $\mb{x}_i$ denotes the extracted feature vector from HFLU for news articles, $\mb{z}_i$ represents the input from other GDUs corresponding to subjects, and $\mb{t}_i$ represents the input from other GDUs about creators. Considering that the GDU for each news article may be connected to multiple GDUs of subjects and creators, the $mean(\cdot)$ of the outputs from the GDUs corresponding to these subjects and creators will be computed as the inputs $\mb{z}_i$ and $\mb{t}_i$ instead respectively, which is also indicated by the GDU architecture illustrated in Figure~\ref{fig:unit}. For the inputs from the subjects, GDU has a gate called the ``forget gate'', which may update some content of $\mb{z}_i$ to forget. The forget gate is important, since in the real world, different news articles may focus on different aspects about the subjects and ``forgetting'' part of the input from the subjects is necessary in modeling. Formally, we can represent the ``forget gate'' together with the updated input as
\begin{equation*}
\tilde{\mb{z}}{i} = \mb{f}_i \otimes \mb{z}_i, \mbox{ where  }\mb{f}_i = \sigma \left( \mb{W}_f \left[\mb{x}_i^\top, \mb{z}_i^\top, \mb{t}_i^\top \right]^\top \right).
\end{equation*}
Here, operator $\otimes$ denotes the entry-wise product of vectors and $\mb{W}_f$ represents the variable of the forget gate in GDU.

Meanwhile, for the input from the creator nodes, a new node-type ``adjust gate'' is introduced in GDU. Here, the term ``adjust'' models the necessary changes of information between different node categories (e.g., from creators to articles). Formally, we can represent the ``adjust gate'' as well as the updated input as
\begin{align*}
\tilde{\mb{t}}_i = \mb{e}_i \otimes \mb{t}_i, \mbox{ where  }\mb{e}_i = \sigma \left( \mb{W}_e \left[\mb{x}_i^\top, \mb{z}_i^\top, \mb{t}_i^\top \right]^\top \right),
\end{align*}
where $\mb{W}_e$ denotes the variable matrix in the adjust gate.

GDU allows different combinations of these input/state vectors, which are controlled by the selection gates $\mb{g}_i$ and $\mb{r}_i$ respectively. Formally, we can represent the final output of GDU as
\begin{align*}
\mb{h}_i &= \mb{g}_i \otimes \mb{r}_i \otimes \tanh \left(\mb{W}_u [\mb{x}_i^\top, \tilde{\mb{z}}_{i}^\top, \tilde{\mb{t}}_i^\top]^\top \right)\\
&\oplus (\mb{1} \ominus \mb{g}_i) \otimes \mb{r}_i \otimes \tanh \left(\mb{W}_u [\mb{x}_i^\top, {\mb{z}}_i^\top, \tilde{\mb{t}}_i^\top]^\top \right)\\
&\oplus \mb{g}_i \otimes (\mb{1} \ominus \mb{r}_i) \otimes \tanh \left(\mb{W}_u [\mb{x}_i^\top, \tilde{\mb{z}}_i^\top, {\mb{t}}_i^\top]^\top \right)\\
&\oplus (\mb{1} \ominus \mb{g}_i) \otimes (\mb{1} \ominus \mb{r}_i) \otimes \tanh \left(\mb{W}_u [\mb{x}_i^\top, {\mb{z}}_i^\top, {\mb{t}}_i^\top]^\top \right),
\end{align*}
where $\mb{g}_i = \sigma ( \mb{W}_g \left[\mb{x}_i^\top, \mb{z}_i^\top, \mb{t}_i^\top \right]^\top )$, and $\mb{r}_i = \sigma ( \mb{W}_r \left[\mb{x}_i^\top, \mb{z}_i^\top, \mb{t}_i^\top \right]^\top )$, and term $\mb{1}$ denotes a vector filled with value $1$. Operators $\oplus$ and $\ominus$ denote the entry-wise addition and minus operation of vectors. Matrices $\mb{W}_u$, $\mb{W}_g$, $\mb{W}_r$ represent the variables involved in the components. Vector $\mb{h}_i$ will be the output of the GDU model.

The introduced GDU model also works for both the news subjects and creator nodes in the network. When applying the GDU to model the states of the subject/creator nodes with two input only, the remaining input port can be assigned with a default value (usually vector $\mb{0}$). Based on the GDU, we can denote the overall architecture of the {\our} as shown in Figure~\ref{fig:architecture}, where the lines connecting the GDUs denote the data flow among the unit models. In the following section, we will introduce how to learn the parameters involved in the {\our} model for concurrent credibility inference of multiple nodes.

\subsection{Deep Diffusive Network Model Learning}

\begin{figure}[t]
\centering
    \begin{minipage}[l]{0.9\columnwidth}
      \centering
      \includegraphics[width=1.0\textwidth]{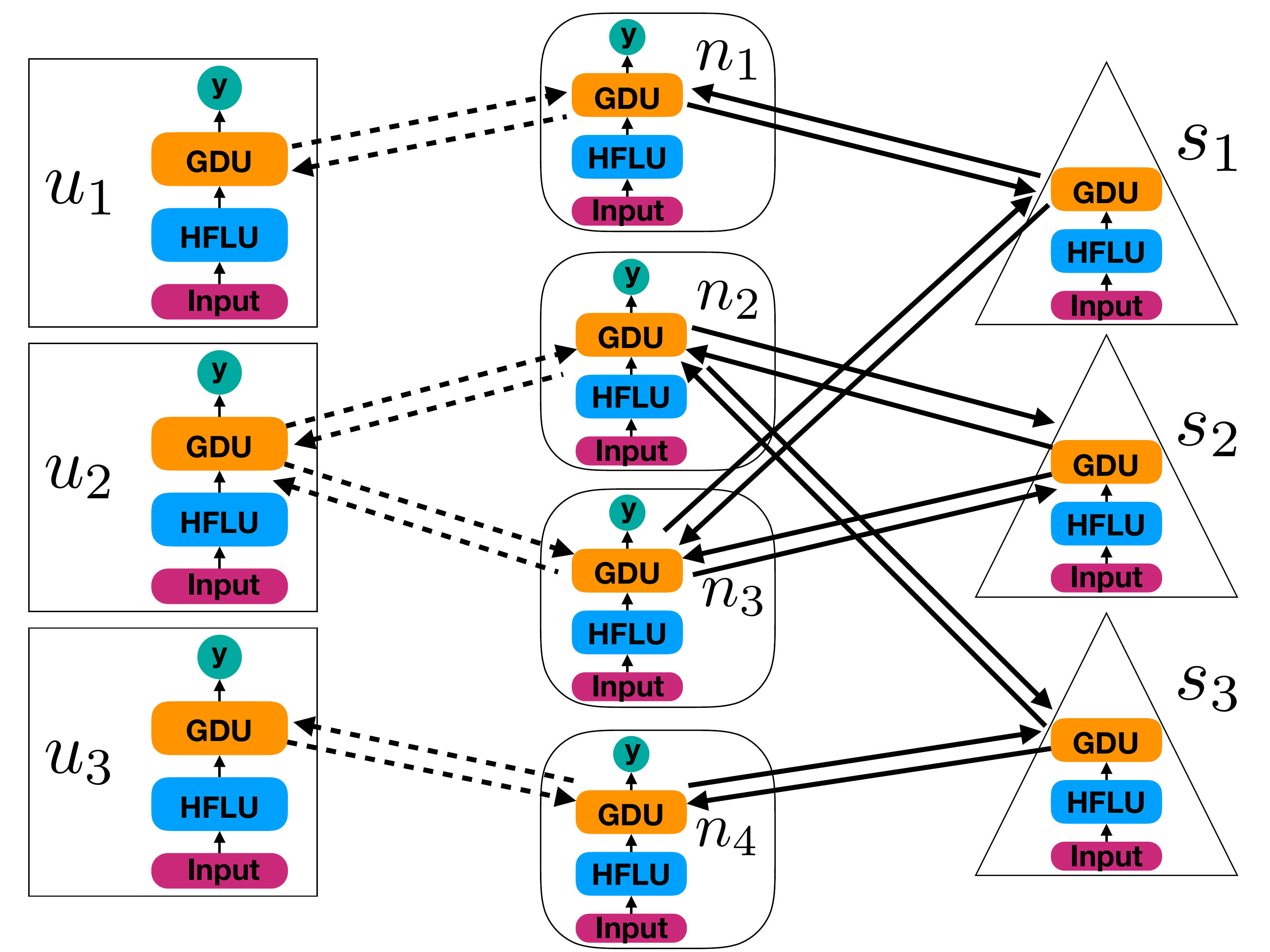}
    \end{minipage}
\caption{The Architecture of Framework {\our}.}\label{fig:architecture}
\end{figure}

In the {\our} model as shown in Figure~\ref{fig:architecture}, based on the output state vectors of news articles, news creators and news subjects, the framework will project the feature vectors to their credibility labels. Formally, given the state vectors $\mb{h}_{n,i}$ of news article $n_i$, $\mb{h}_{u,j}$ of news creator $u_j$, and $\mb{h}_{s,l}$ of news subject $s_l$, we can represent their inferred credibility labels as vectors $\mb{y}_{n,i}, \mb{y}_{u,j}, \mb{y}_{s,l} \in \mathcal{R}^{|\mathcal{Y}|}$ respectively, which can be represented as
$$\begin{cases}
\mb{y}_{n,i} &= softmax\left( \mb{W}_n \mb{h}_{n,i} \right),\\
\mb{y}_{u,j} &= softmax\left( \mb{W}_u \mb{h}_{u,j} \right),\\
\mb{y}_{s,l} &= softmax\left( \mb{W}_s \mb{h}_{s,l} \right).
\end{cases}$$
where $\mb{W}_u$, $\mb{W}_n$ and $\mb{W}_s$ define the weight variables projecting state vectors to the output vectors, and function $softmax(\cdot)$ represents the softmax function.

Meanwhile, based on the news articles in the training set $\mathcal{T}_n \subset \mathcal{N}$ with the ground-truth credibility label vectors $\{\hat{\mb{y}}_{n,i}\}_{n_i \in \mathcal{T}_n}$, we can define the loss function of the framework for news article credibility label learning as the cross-entropy between the prediction results and the ground truth:
\begin{align*}
\mathcal{L}(\mathcal{T}_n) &= - \sum_{n_i \in \mathcal{T}_n}  \sum_{k=1}^{|\mathcal{Y}|}\hat{\mb{y}}_{n,i}[k] \log {\mb{y}}_{n,i}[k].
\end{align*}
Similarly, we can define the loss terms introduced by news creators and subjects based on training sets $\mathcal{T}_u \subset \mathcal{U}$ and $\mathcal{T}_s \subset \mathcal{S}$ as
\begin{align*}
\mathcal{L}(\mathcal{T}_u) &= - \sum_{u_j \in \mathcal{T}_u}  \sum_{k=1}^{|\mathcal{Y}|}\hat{\mb{y}}_{u,j}[k] \log {\mb{y}}_{u,j}[k],
\end{align*}
\begin{align*}
\mathcal{L}(\mathcal{T}_s) &= - \sum_{s_l \in \mathcal{T}_s}  \sum_{k=1}^{|\mathcal{Y}|}\hat{\mb{y}}_{s,l}[k] \log {\mb{y}}_{s,l}[k],
\end{align*}
where $\mb{y}_{u,j}$ and $\hat{\mb{y}}_{u,j}$ (and $\mb{y}_{s,l}$ and $\hat{\mb{y}}_{s,l}$) denote the prediction result vector and ground-truth vector of creator (and subject) respectively.

Formally, the main objective function of the {\our} model can be represented as follows:
\begin{equation*}
\min_{\mb{W}} \mathcal{L}(\mathcal{T}_n) + \mathcal{L}(\mathcal{T}_u) + \mathcal{L}(\mathcal{T}_s) + \alpha \cdot \mathcal{L}_{reg}(\mb{W}),
\end{equation*}
where $\mb{W}$ denotes all the involved variables to be learned, term $\mathcal{L}_{reg}(\mb{W})$ represents the regularization term (i.e., the sum of $L_2$ norm on the variable vectors and matrices), and $\alpha$ denotes the regularization term weight. By resolving the optimization functions, we will be able to learn the variables involved in the framework. In this paper, we propose to train the framework with the \textit{back-propagation} algorithm. For the news articles, creators and subjects in the testing set, their predicted credibility labels will be outputted as the final result.

\section{Experiments}\label{sec:experiment}

To test the effectiveness of the proposed model, in this part, extensive experiments will be done on the real-world fake news dataset, PolitiFact, which has been analyzed in Section~\ref{sec:data_analysis} in great detail. The dataset statistical information is also available in Table~\ref{tab:datastat}. In this section, we will first introduce the experimental settings, and then provide experimental results together with the detailed analysis.

\subsection{Experimental Settings}

The experimental setting covers (1) detailed experimental setups, (2) comparison methods and (3) evaluation metrics, which will be introduced as follows respectively.

\subsubsection{Experimenta Setups}

Based on the input PolitiFact dataset, we can represent the set of news articles, creators and subjects as $\mathcal{N}$, $\mathcal{U}$ and $\mathcal{S}$ respectively. With 10-fold cross validation, we propose to partition the news article, creator and subject sets into two subsets according to ratio $9:1$ respectively, where $9$ folds are used as the training sets and $1$ fold is used as the testing sets. Here, to simulate the cases with different number of training data. We further sample a subset of news articles, creators and subjects from the training sets, which is controlled by the sampling ratio parameter $\theta \in \{0.1, 0.2, \cdots, 1.0\}$. Here, $\theta = 0.1$ denotes $10\%$ of instances in the $9$ folds are used as the final training set, and $\theta = 1.0$ denotes $100\%$ of instances in the $9$ folds are used as the final training set. The known news article credibility labels will be used as the ground truth for model training and evaluation. Furthermore, based on the categorical labels of news articles, we propose to represent the $6$ credibility labels with $6$ numerical scores instead, and the corresponding relationships are as follows: ``True'': 6, ``Mostly True'': 5, ``Half True'': 4, ``Mostly False'': 3, ``False'': 2, ``Pants on Fire!'': 1. According to the known creator-article and subject-article relationships, we can also compute the credibility scores of creators and subjects, which can be denoted as the weighted sum of credibility scores of published articles (here, the weight denotes the percentage of articles in each class). And the credibility labels corresponding to the creator/subject round scores will be used as the ground truth as well. Based on the training sets of news articles, creators and subjects, we propose to build the {\our} model with their known textual contents, article-creator relationships, and article-subject relationships, and further apply the learned {\our} model to the test sets.

\subsubsection{Comparison Methods}

In the experiments, we compare {\our} extensively with many baseline methods. The list of used comparison methods are listed as follows:
\begin{itemize}
\item \textit{{\our}}: Framework {\our} proposed in this paper can infer the credibility labels of news articles, creators and subjects with both explicit and latent textual features and relationship connections based on the {\unit} model.
\vspace{5pt}
\item \textit{\cnn}: Model {\cnn} introduced in \cite{W17} can effectively classify the textual news articles based on the convolutional neural network (CNN) model.
\vspace{5pt}
\item \textit{\liwc}: In \cite{RCJVC17}, a set of LIWC (Linguistic Inquiry and Word Count) based language features have been extracted, which together with the textual content information can be used in some models, e.g., LSTM as used in \cite{RCJVC17}, for news article classification.
\vspace{5pt}
\item \textit{\trifn}: The {\trifn} model proposed in \cite{SWL17} exploits the user, news and publisher relationships for fake news detection. In the experiments, we also extend this model by replacing the publisher with subjects and compare with the other methods as a baseline method.
\vspace{5pt}
\item \textit{{\deepwalk}}: Model {\deepwalk} \cite{PAS14} is a network embedding model. Based on the fake news network structure, {\deepwalk} embeds the news articles, creators and subjects to a latent feature space. Based on the learned embedding results, we can further build a SVM model to determine the class labels of the new articles, creators and subjects.
\vspace{5pt}
\item \textit{{\modelline}}: The {\modelline} model is a scalable network embedding model proposed in \cite{TQWZYM15}, which optimizes an objective function that preserves both the local and global network structures. Similar to {\deepwalk}, based on the embedding results, a classifier model can be further build to classify the news articles, creators and subjects.
\vspace{5pt}
\item \textit{{\propagation}}: In addition, merely based on the fake news heterogeneous network structure, we also propose to compare the above methods with a label-propagation based model proposed in \cite{HXZZGY16}, which also considers the node types and link types into consideration. The prediction score will be rounded and cast into labels according to the label-score mappings.
\vspace{5pt}
\item \textit{{\rnn}}: In this paper, merely based on the textual contents, we propose to apply the {\rnn} model \cite{MKBCK10} to learn the latent representations of the textual input. Furthermore, the latent feature vectors will be fused to predict the news article, creator and subject credibility labels.
\vspace{5pt}
\item \textit{{\svm}}: Slightly different form {\rnn}, based on the raw text inputs, a set of explicit features can be extracted according to the descriptions in this paper, which will be used as the input for building a {\svm} \cite{libsvm} based classification model as the last baseline method.
\end{itemize}

Here, we need to add a remark, according to \cite{W17, RCJVC17, SWL17}, methods {\cnn}, {\liwc} and {\trifn} are proposed for classifying the news articles only, which will not be applied for the label inference of subjects and creators in the experiments. Among the other baseline methods, {\deepwalk} and {\modelline} use the network structure information only, and all build a classification model based on the network embedding results. Model {\propagation} also only uses the network structure information, but is based on the label propagation model instead. Both {\rnn} and {\svm} merely utilize the textual contents only, but their differences lies in:  {\rnn} builds the classification model based on the latent features and {\svm} builds the classification model based on the explicit features instead.



\begin{figure*}[t]
\centering
\subfigure[Bi-Class Article Accuracy]{ \label{fig:bi_article_acc}
    \begin{minipage}[l]{0.55\columnwidth}
      \centering
      \includegraphics[width=1.1\textwidth]{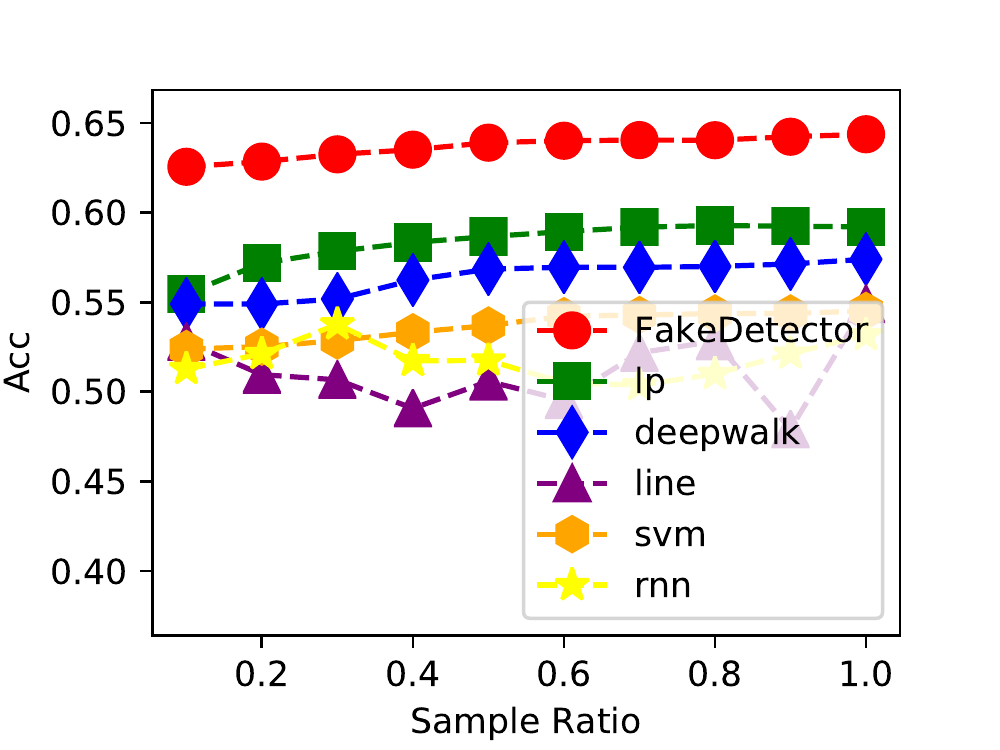}
    \end{minipage}
}
\subfigure[Bi-Class Article F1]{\label{fig:bi_article_f1}
    \begin{minipage}[l]{0.55\columnwidth}
      \centering
      \includegraphics[width=1.1\textwidth]{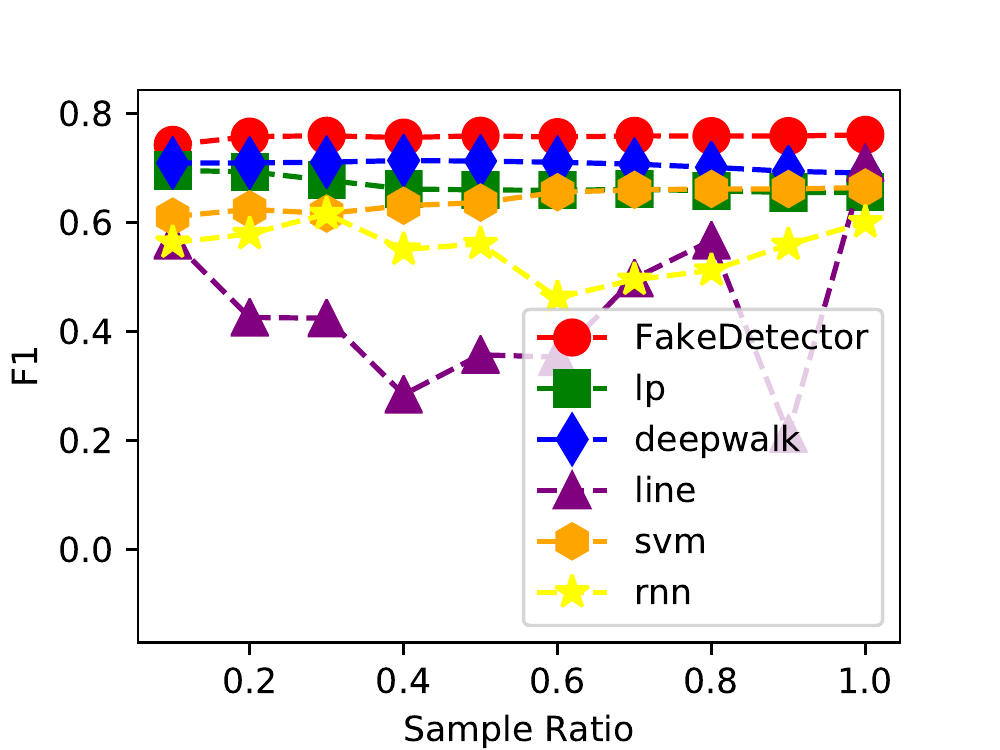}
    \end{minipage}
}
\subfigure[Bi-Class Article Precision]{ \label{fig:bi_article_prec}
    \begin{minipage}[l]{0.55\columnwidth}
      \centering
      \includegraphics[width=1.1\textwidth]{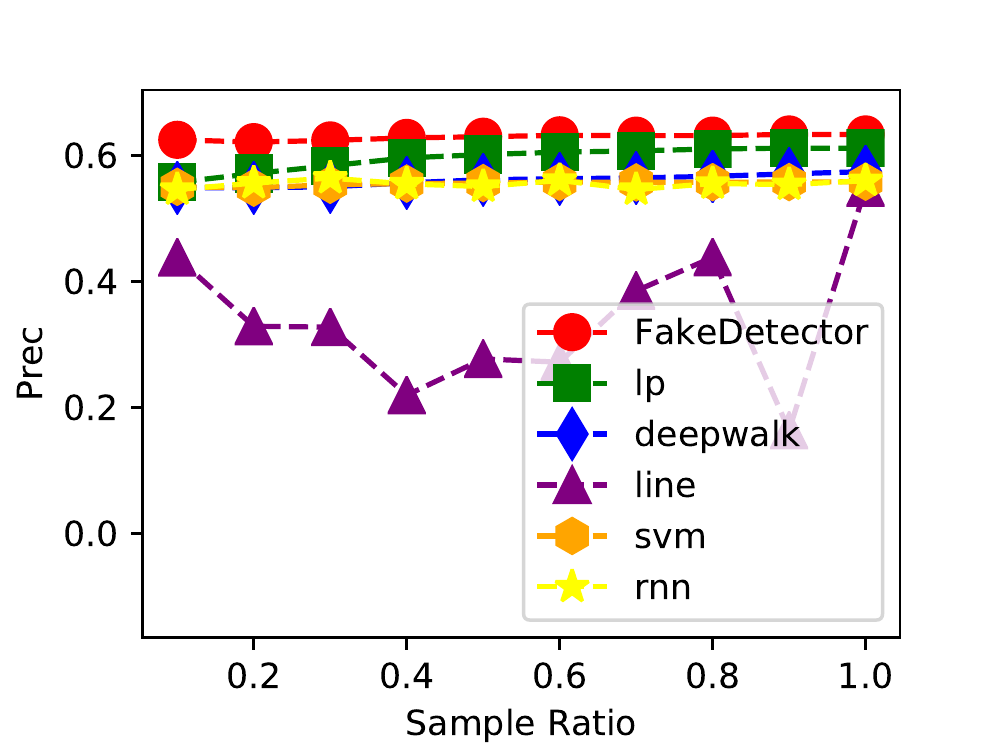}
    \end{minipage}
}
\subfigure[Bi-Class Article Recall]{ \label{fig:bi_article_recall}
    \begin{minipage}[l]{0.55\columnwidth}
      \centering
      \includegraphics[width=1.1\textwidth]{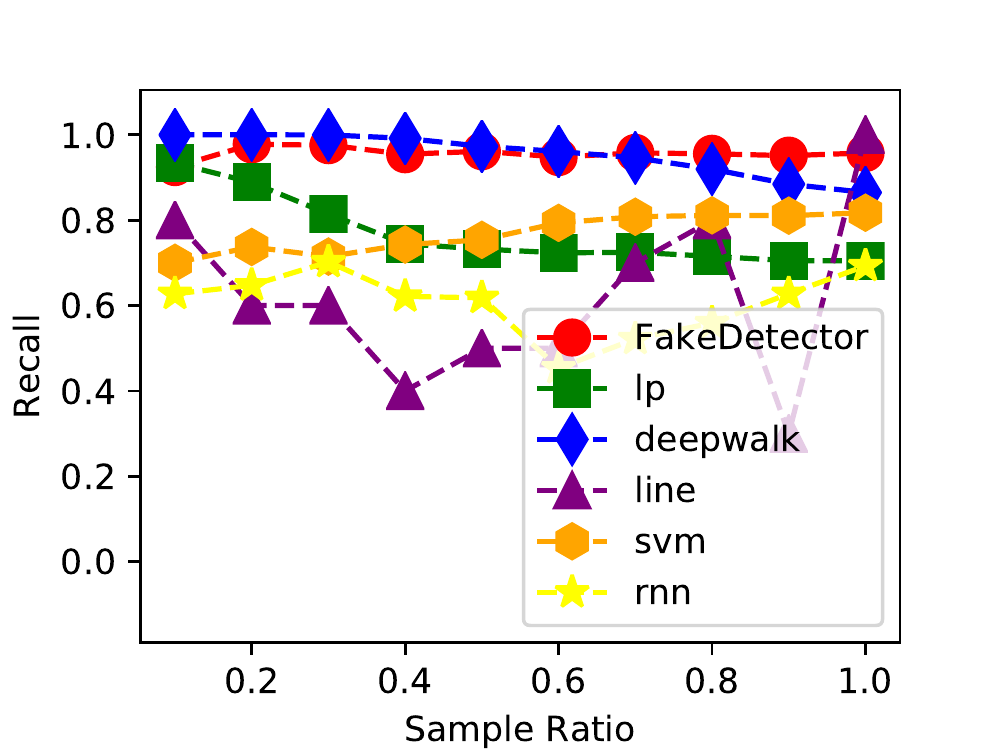}
    \end{minipage}
}
\subfigure[Bi-Class Creator Accuracy]{ \label{fig:bi_creator_acc}
    \begin{minipage}[l]{0.55\columnwidth}
      \centering
      \includegraphics[width=1.1\textwidth]{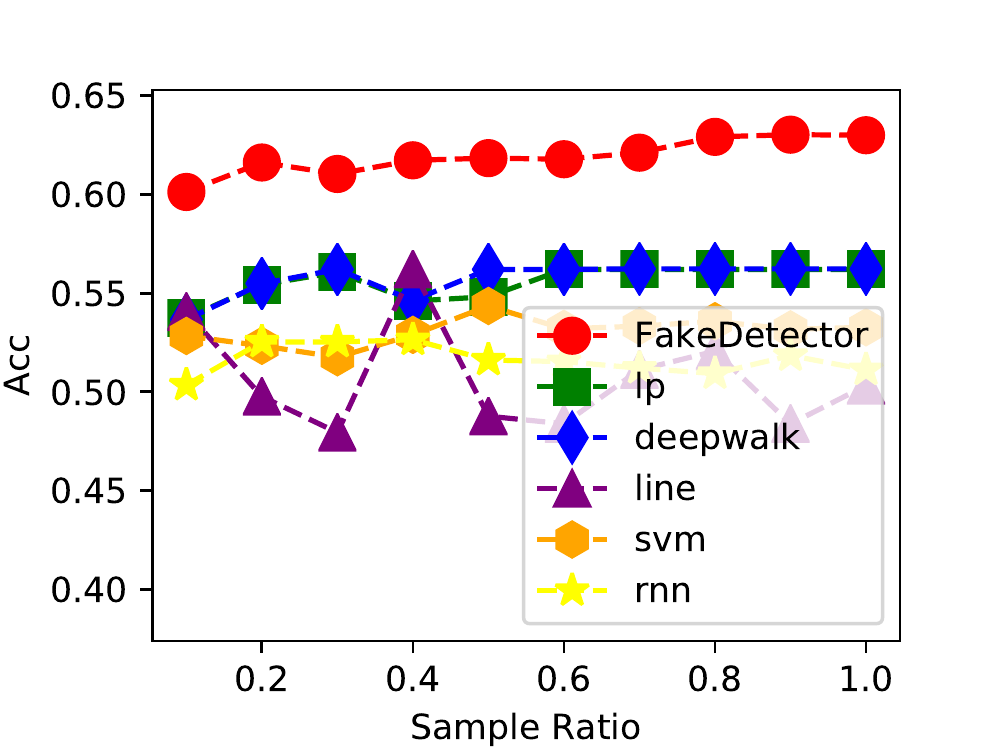}
    \end{minipage}
}
\subfigure[Bi-Class Creator F1]{\label{fig:bi_creator_f1}
    \begin{minipage}[l]{0.55\columnwidth}
      \centering
      \includegraphics[width=1.1\textwidth]{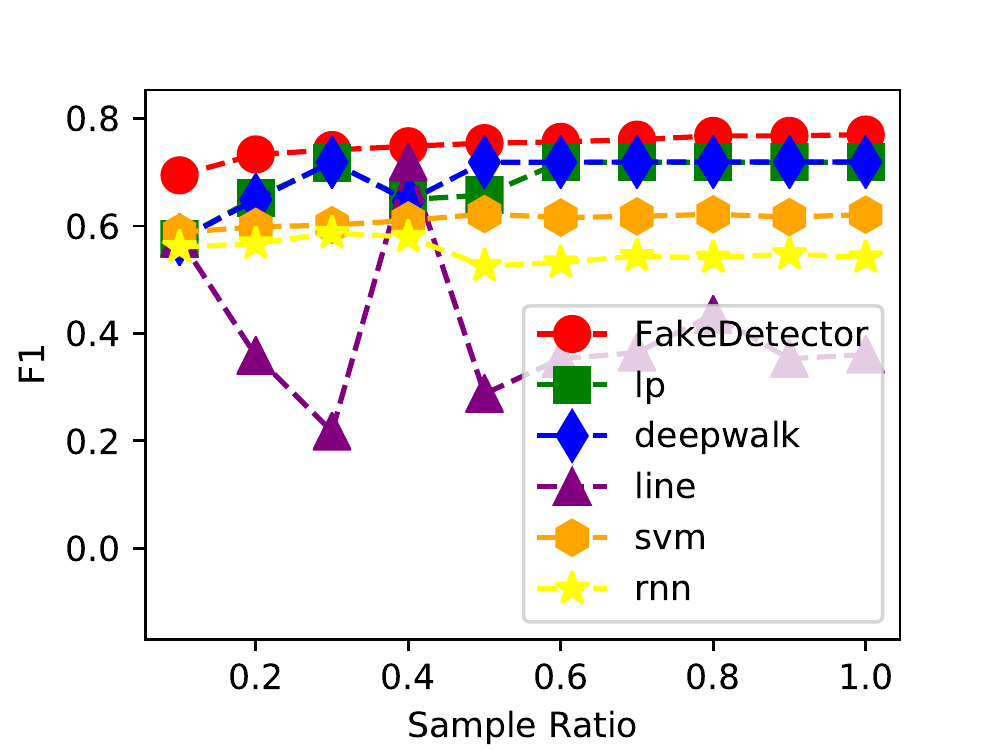}
    \end{minipage}
}
\subfigure[Bi-Class Creator Precision]{ \label{fig:bi_creator_prec}
    \begin{minipage}[l]{0.55\columnwidth}
      \centering
      \includegraphics[width=1.1\textwidth]{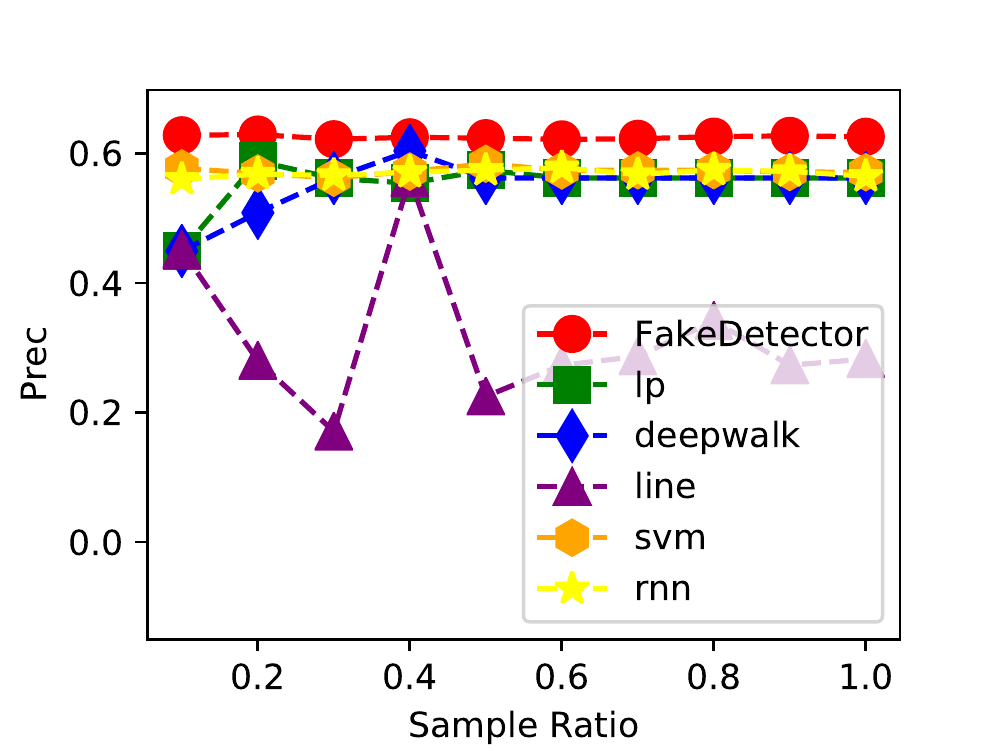}
    \end{minipage}
}
\subfigure[Bi-Class Creator Recall]{ \label{fig:bi_creator_recall}
    \begin{minipage}[l]{0.55\columnwidth}
      \centering
      \includegraphics[width=1.1\textwidth]{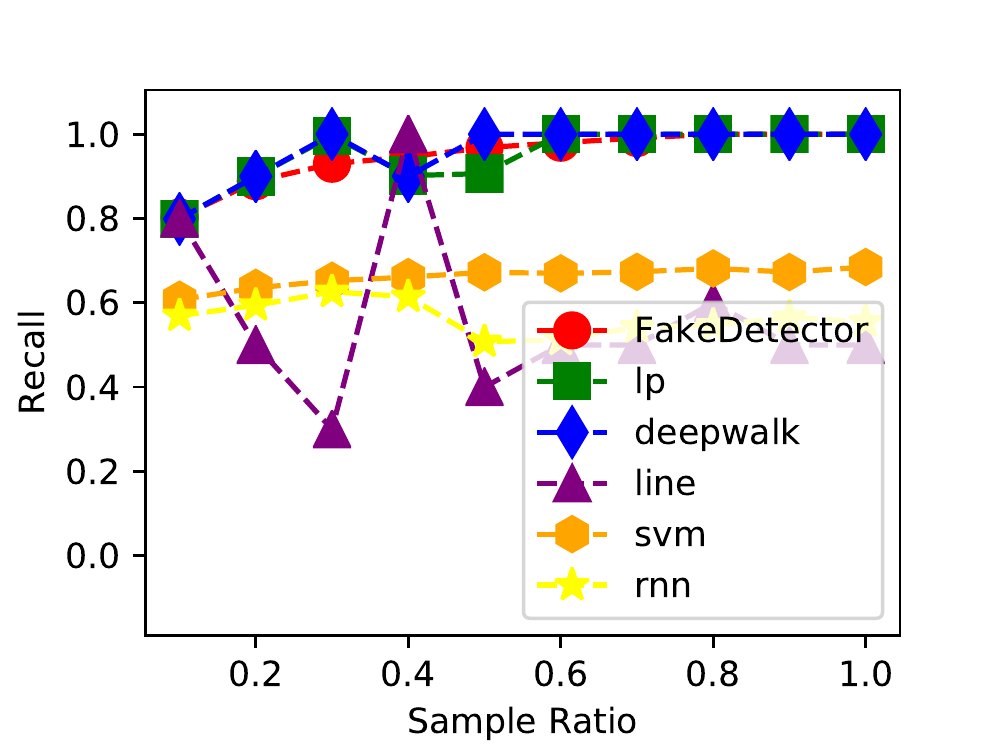}
    \end{minipage}
}
\subfigure[Bi-Class Subject Accuracy]{ \label{fig:bi_subject_acc}
    \begin{minipage}[l]{0.55\columnwidth}
      \centering
      \includegraphics[width=1.1\textwidth]{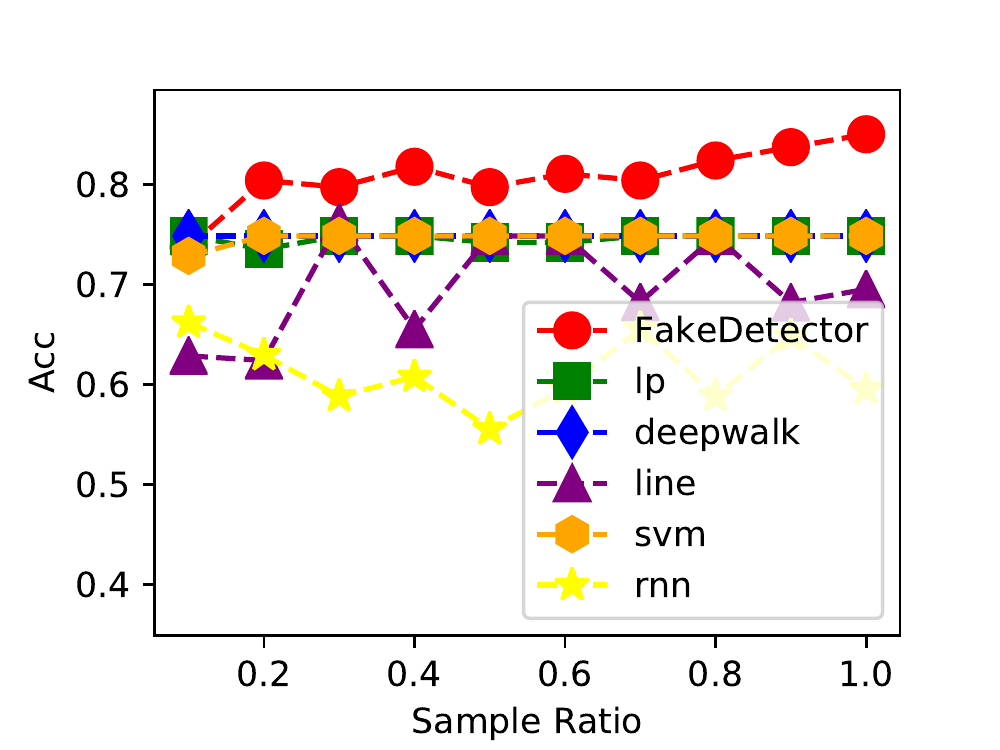}
    \end{minipage}
}
\subfigure[Bi-Class Subject F1]{\label{fig:bi_subject_f1}
    \begin{minipage}[l]{0.55\columnwidth}
      \centering
      \includegraphics[width=1.1\textwidth]{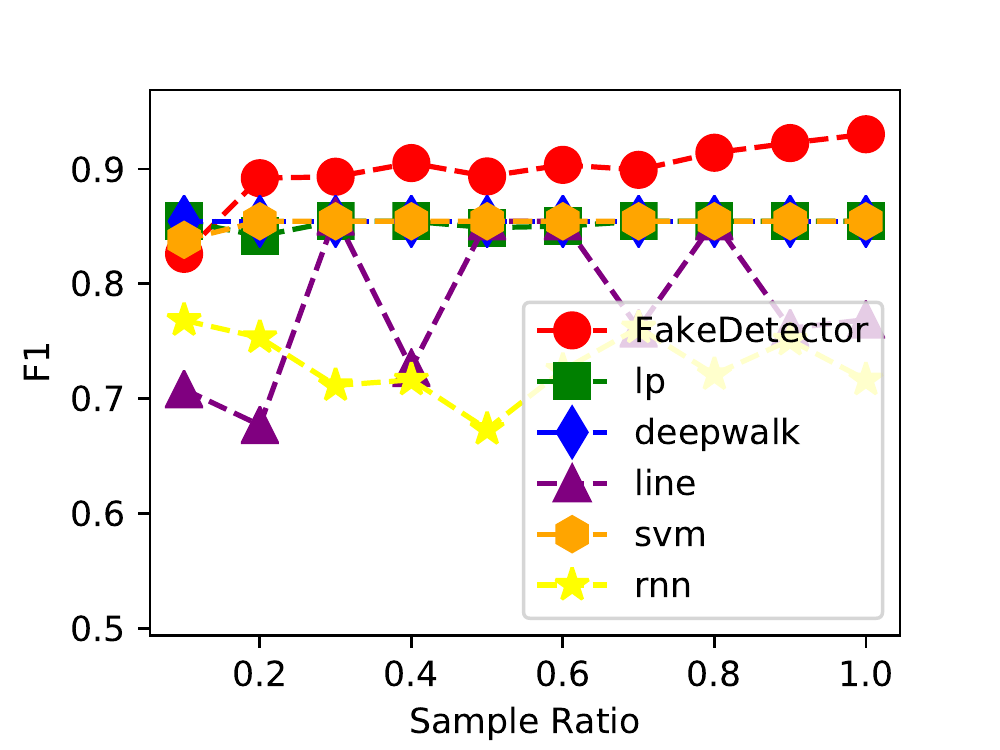}
    \end{minipage}
}
\subfigure[Bi-Class Subject Precision]{ \label{fig:bi_subject_prec}
    \begin{minipage}[l]{0.55\columnwidth}
      \centering
      \includegraphics[width=1.1\textwidth]{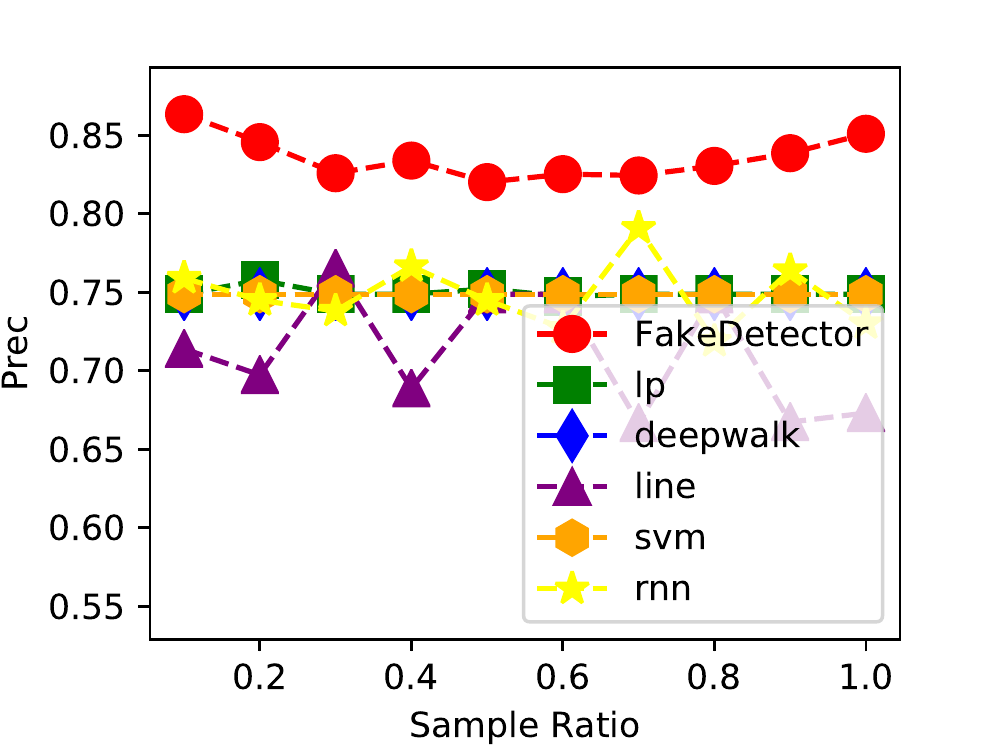}
    \end{minipage}
}
\subfigure[Bi-Class Subject Recall]{ \label{fig:bi_subject_recall}
    \begin{minipage}[l]{0.55\columnwidth}
      \centering
      \includegraphics[width=1.1\textwidth]{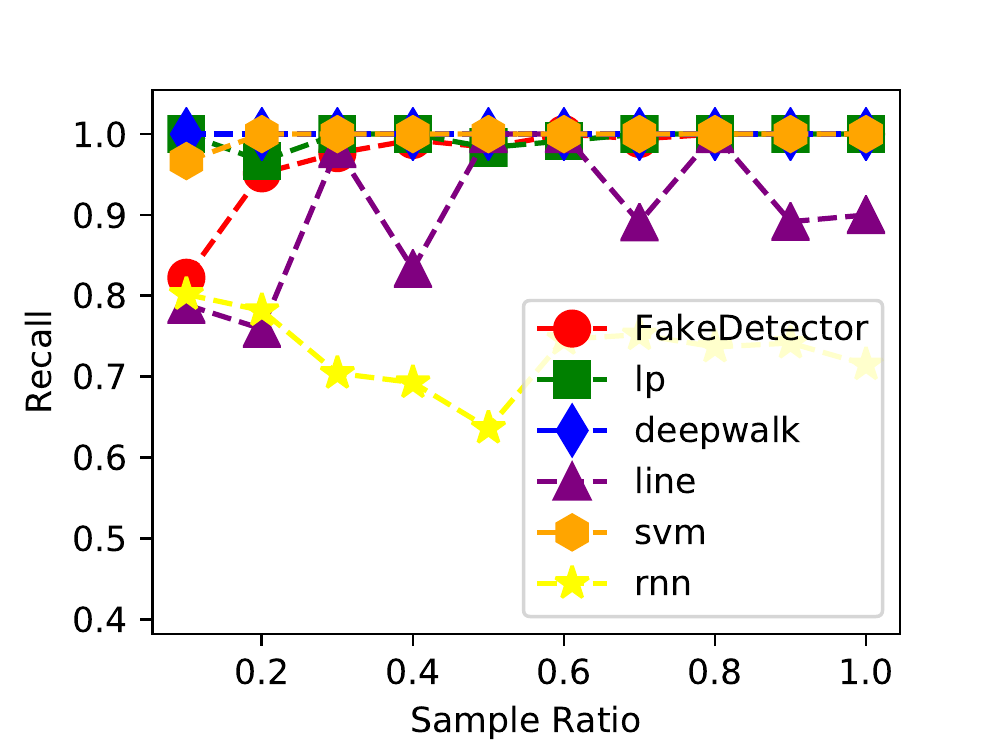}
    \end{minipage}
}
\caption{Bi-Class Credibility Inference of News Articles \ref{fig:bi_article_acc}-\ref{fig:bi_article_recall}, Creators \ref{fig:bi_creator_acc}-\ref{fig:bi_creator_recall} and Subjects \ref{fig:bi_subject_acc}-\ref{fig:bi_subject_recall}.}\label{fig:bi_result}
\end{figure*}

\subsubsection{Evaluation Metrics}

Several frequently-used classification evaluation metrics will be used for the performance evaluation of the comparison methods. In the evaluation, we will cast the credibility inference problem into a binary class classification and a multi-class classification problem respectively. By grouping class labels \{True, Mostly True, Half True\} as the positive class and labels \{Pants on Fire!, False, Mostly False\} as the negative class, the credibility inference problem will be modeled as a binary-class classification problem, whose results can be evaluated by metrics, like Accuracy, Precision, Recall and F1. Meanwhile, if the model infers the original $6$ class labels \{True, Mostly True, Half True, Mostly False, False, Pants on Fire!\} directly, the problem will be a multi-class classification problem, whose performance can be evaluated by metrics, like Accuracy, Macro Precision, Macro Recall and Macro F1 respectively.

\begin{figure*}[t]
\centering
\subfigure[Multi-Class Article Accuracy]{ \label{fig:multi_article_acc}
    \begin{minipage}[l]{0.55\columnwidth}
      \centering
      \includegraphics[width=1.1\textwidth]{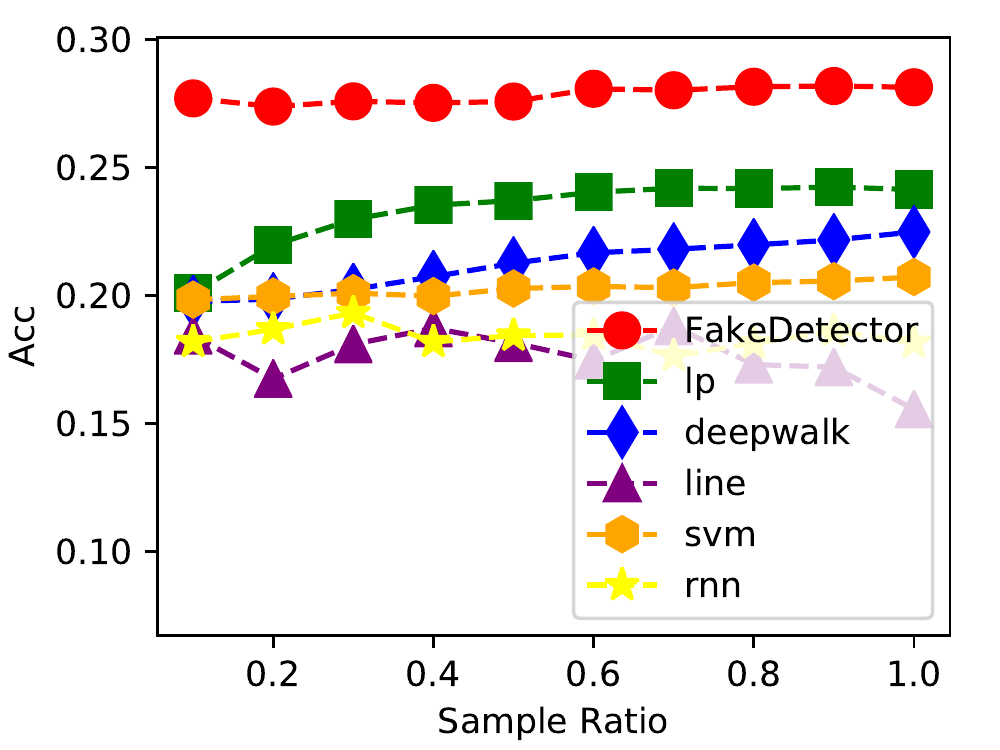}
    \end{minipage}
}
\subfigure[Multi-Class Article F1]{\label{fig:multi_article_f1}
    \begin{minipage}[l]{0.55\columnwidth}
      \centering
      \includegraphics[width=1.1\textwidth]{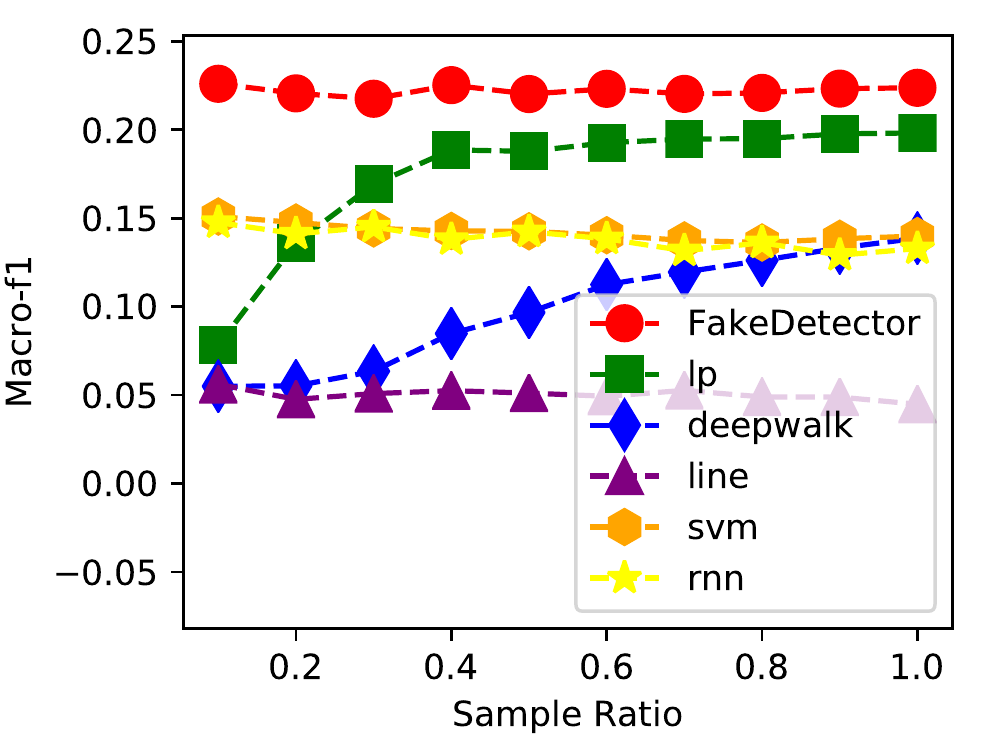}
    \end{minipage}
}
\subfigure[Multi-Class Article Precision]{ \label{fig:multi_article_prec}
    \begin{minipage}[l]{0.55\columnwidth}
      \centering
      \includegraphics[width=1.1\textwidth]{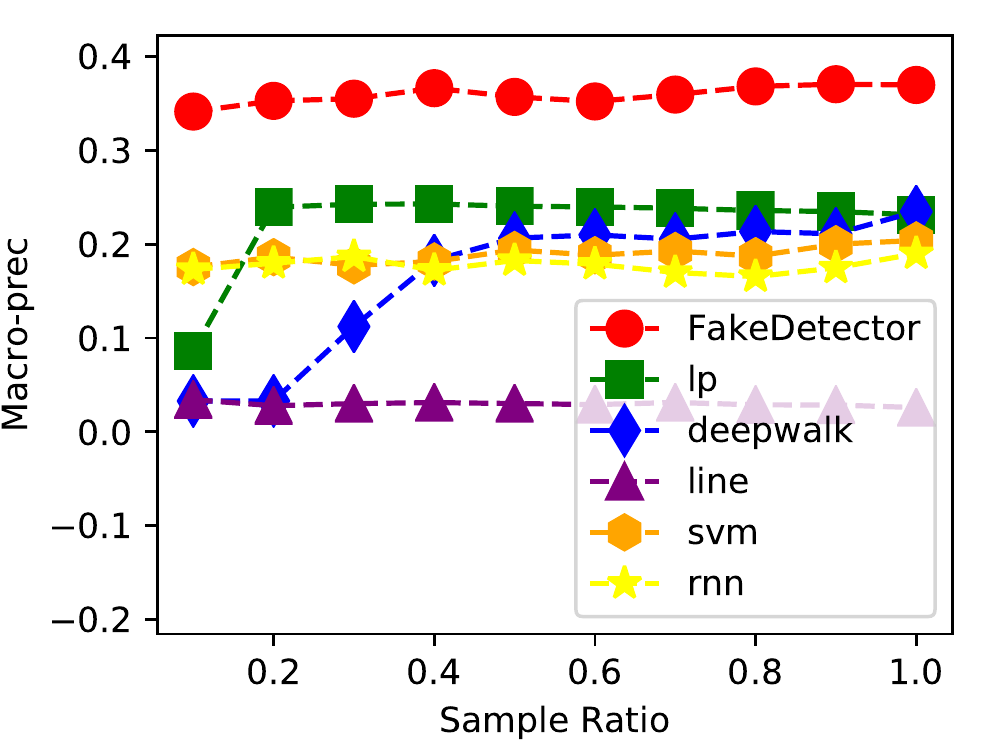}
    \end{minipage}
}
\subfigure[Multi-Class Article Recall]{ \label{fig:multi_article_recall}
    \begin{minipage}[l]{0.55\columnwidth}
      \centering
      \includegraphics[width=1.1\textwidth]{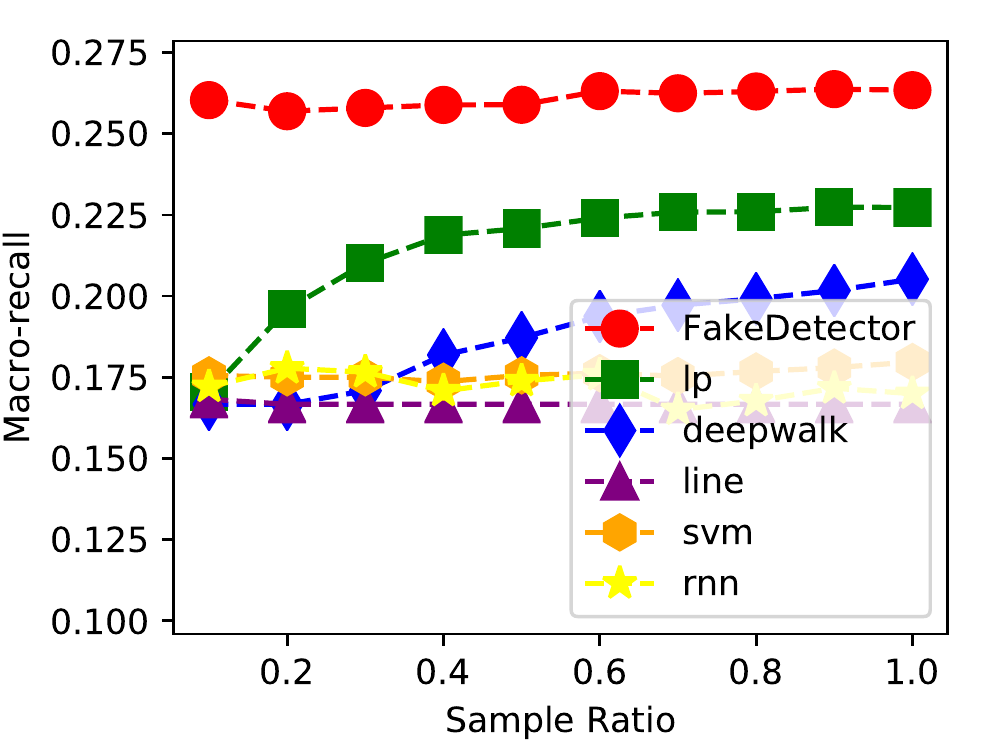}
    \end{minipage}
}
\subfigure[Multi-Class Creator Accuracy]{ \label{fig:multi_creator_acc}
    \begin{minipage}[l]{0.55\columnwidth}
      \centering
      \includegraphics[width=1.1\textwidth]{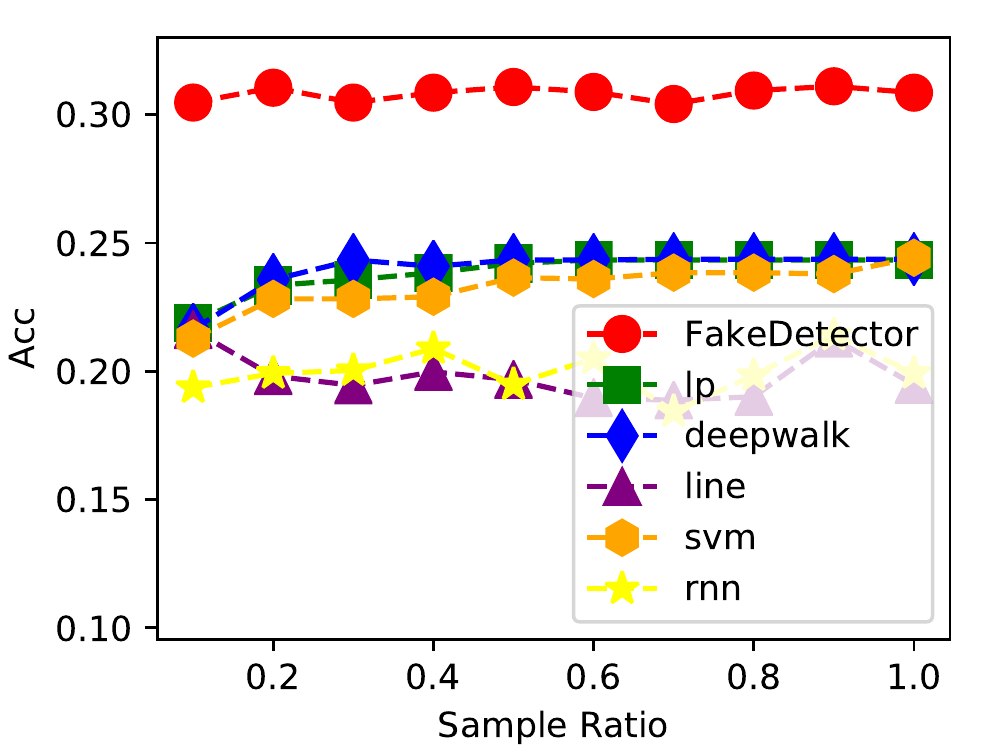}
    \end{minipage}
}
\subfigure[Multi-Class Creator F1]{\label{fig:multi_creator_f1}
    \begin{minipage}[l]{0.55\columnwidth}
      \centering
      \includegraphics[width=1.1\textwidth]{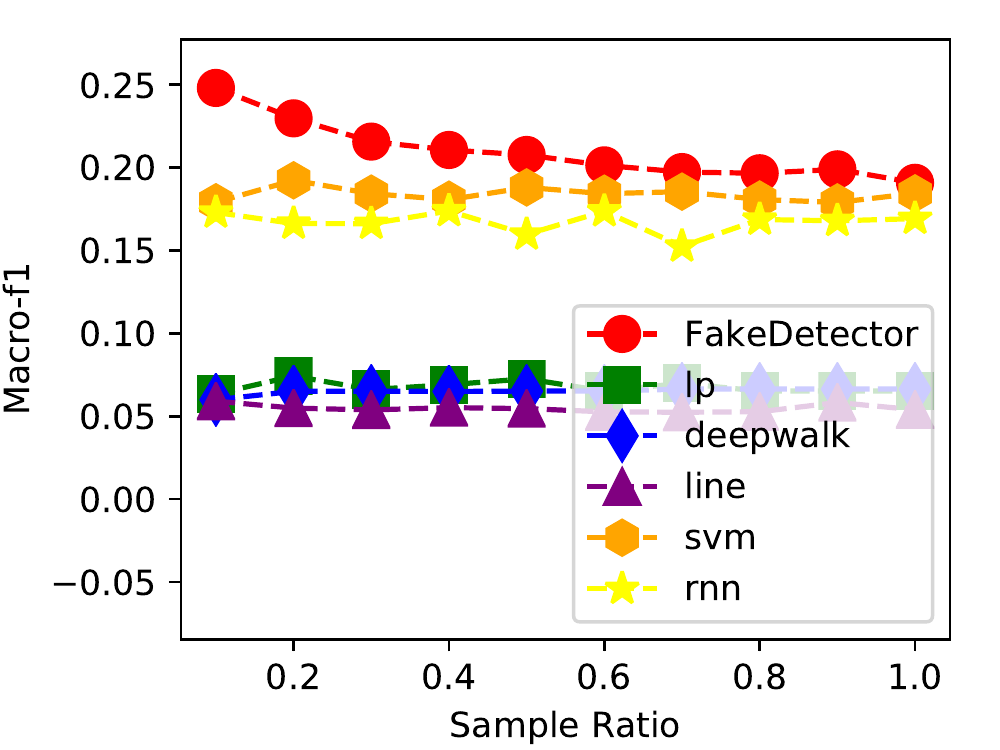}
    \end{minipage}
}
\subfigure[Multi-Class Creator Precision]{ \label{fig:multi_creator_prec}
    \begin{minipage}[l]{0.55\columnwidth}
      \centering
      \includegraphics[width=1.1\textwidth]{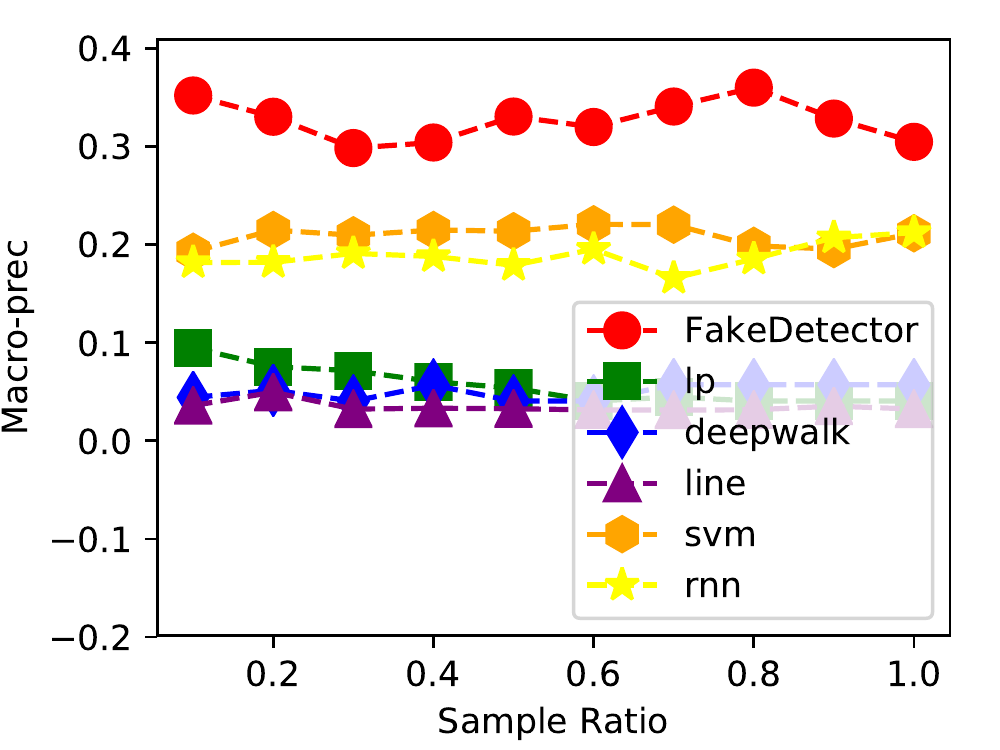}
    \end{minipage}
}
\subfigure[Multi-Class Creator Recall]{ \label{fig:multi_creator_recall}
    \begin{minipage}[l]{0.55\columnwidth}
      \centering
      \includegraphics[width=1.1\textwidth]{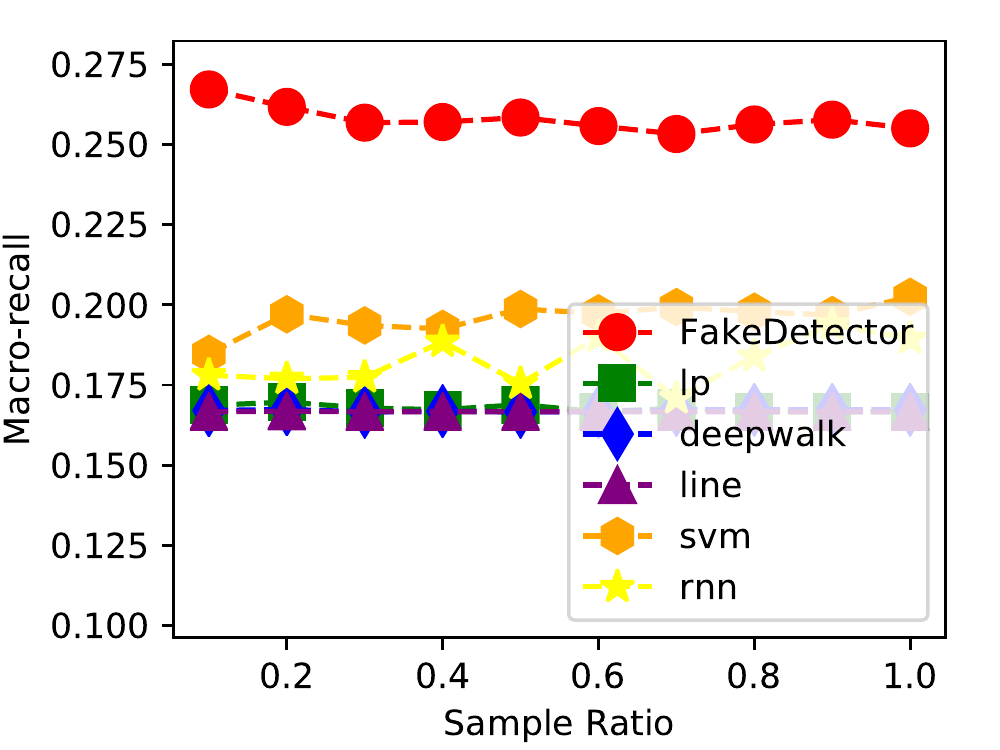}
    \end{minipage}
}
\subfigure[Multi-Class Subject Accuracy]{ \label{fig:multi_subject_acc}
    \begin{minipage}[l]{0.55\columnwidth}
      \centering
      \includegraphics[width=1.1\textwidth]{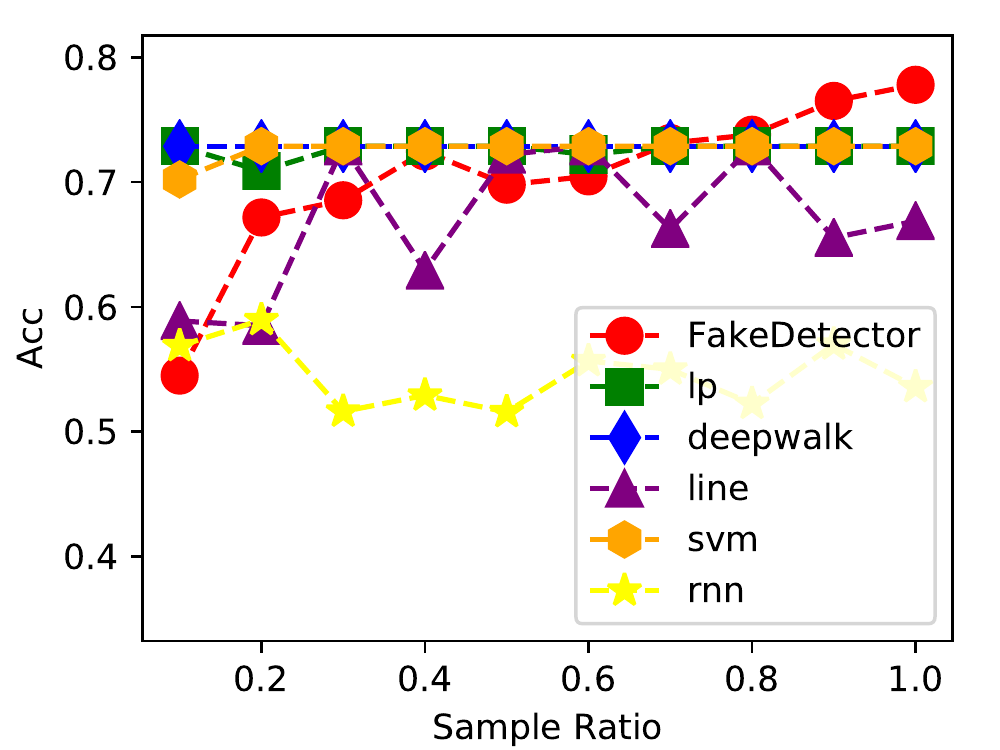}
    \end{minipage}
}
\subfigure[Multi-Class Subject F1]{\label{fig:multi_subject_f1}
    \begin{minipage}[l]{0.55\columnwidth}
      \centering
      \includegraphics[width=1.1\textwidth]{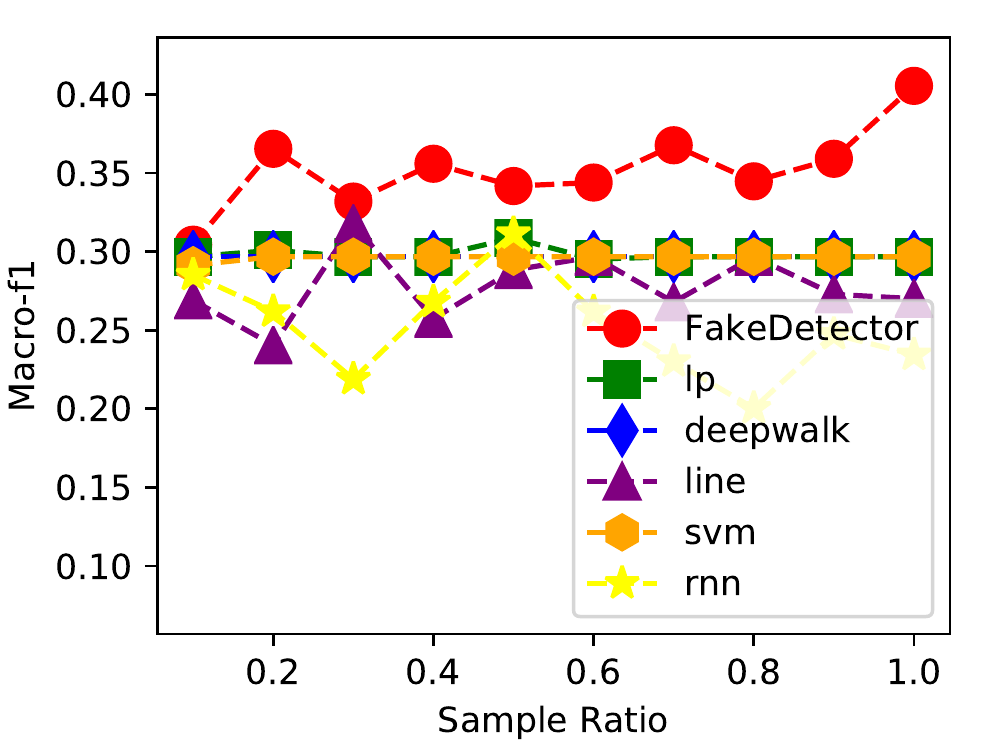}
    \end{minipage}
}
\subfigure[Multi-Class Subject Precision]{ \label{fig:multi_subject_prec}
    \begin{minipage}[l]{0.55\columnwidth}
      \centering
      \includegraphics[width=1.1\textwidth]{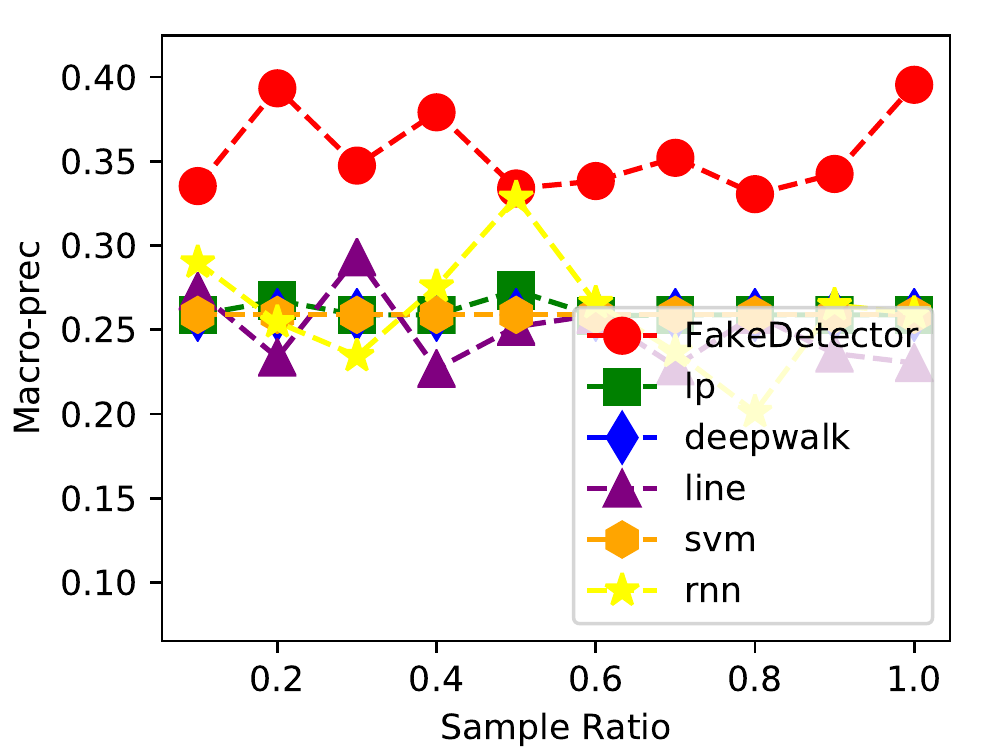}
    \end{minipage}
}
\subfigure[Multi-Class Subject Recall]{ \label{fig:multi_subject_recall}
    \begin{minipage}[l]{0.55\columnwidth}
      \centering
      \includegraphics[width=1.1\textwidth]{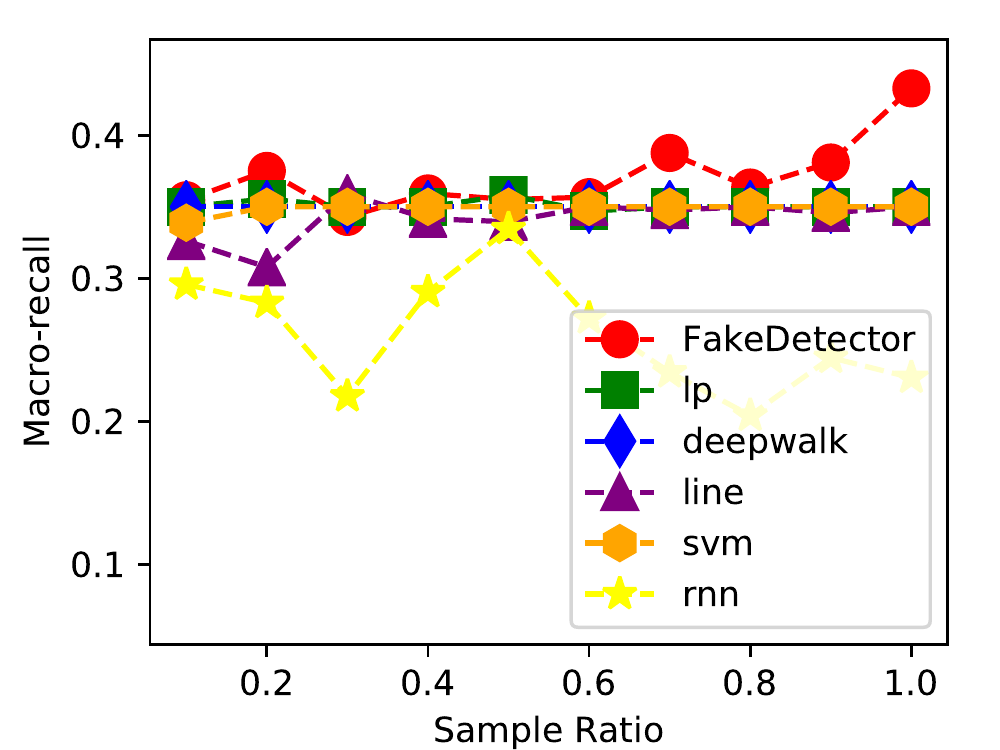}
    \end{minipage}
}
\caption{Multi-Class Credibility Inference of News Articles \ref{fig:multi_article_acc}-\ref{fig:multi_article_recall}, Creators \ref{fig:multi_creator_acc}-\ref{fig:multi_creator_recall} and Subjects \ref{fig:multi_subject_acc}-\ref{fig:multi_subject_recall}.}\label{fig:multi_result}
\end{figure*}

\subsection{Experimental Results}

The experimental results are provided in Figures~\ref{fig:bi_result}-\ref{fig:multi_result}, where the plots in Figure~\ref{fig:bi_result} are about the binary-class inference results of articles, creators and subjects, while the plots in Figure~\ref{fig:multi_result} are about the multi-class inference results.

\subsubsection{Bi-Class Inference Results}

According to the plots in Figure~\ref{fig:bi_result}, method {\our} can achieve the best performance among all the other methods in inferring the bi-class labels of news articles, creators and subjects (for all the evaluation metrics except Recall) with different sample ratios consistently. For instance, when the sample ratio $\theta = 0.1$, the Accuracy score obtained by {\our} in inferring the news articles is $0.63$, which is more than $14.5\%$ higher than the Accuracy score obtained by the state-of-the-art fake news detection models {\cnn}, {\liwc} and {\trifn}, as well as the network structure based models {\propagation}, {\deepwalk}, {\modelline} and the textual content based methods {\rnn} and {\svm}. Similar observations can be identified for the inference of creator credibility and subject credibility respectively. 

Among all the True news articles, creators and subjects identified by {\our}, a large proportion of them are the correct predictions. As shown in the plots, method {\our} can achieve the highest Precision score among all these methods, especially for the subjects. Meanwhile, the Recall obtained by {\our} is slightly lower than the other methods. By studying the prediction results, we observe that {\our} does predict less instances with the ``True'' label, compared with the other methods. The overall performance of {\our} (by balancing Recall and Precision) will surpass the other methods, and the F1 score obtained by {\our} greatly outperforms the other methods.

\subsubsection{Multi-Class Inference Results}

Besides the simplified bi-class inference problem setting, we further infer the information entity credibility at a finer granularity: infer the labels of instances based the original $6$-class label space. The inference results of all the comparison methods are available in Figure~\ref{fig:multi_result}. Generally, according to the performance, the advantages of {\our} are much more significant compared with the other methods in the multi-class prediction setting. For instance, when $\theta = 0.1$, the Accuracy score achieved by {\our} in inferring news article credibility score is $0.28$, which is more than $40\%$ higher than the Accuracy obtained by the other methods. In the multi-class scenario, both the Macro-Precision and Macro-Recall scores of {\our} are also much higher than the other methods. Meanwhile, by comparing the inference scores obtained by the methods in Figures~\ref{fig:bi_result} and Figure~\ref{fig:multi_result}, the multi-class credibility inference scenario is much more difficult and the scores obtained by the methods are much lower than the bi-class inference setting.

%
%
%


\section{Related Work} \label{sec:related_work}

Several research topics are closely correlated with this paper, including \textit{fake news analysis}, \textit{spam detection} and \textit{deep learning}, which will be briefly introduced as follows.\\

\noindent \textbf{Fake News Preliminary Works}: Due the increasingly realized impacts of fake news since the 2016 election, some preliminary research works have been done on fake news detection. The first work on online social network fake news analysis for the election comes from Allcott et al. \cite{AG17}. The other published preliminary works mainly focus on fake news detection instead \cite{RCCC16, SDSRG17, SSWTL17, TBDMA17}. Rubin et al. \cite{RCCC16} provides a conceptual overview to illustrate the unique features of fake news, which tends to mimic the format and style of journalistic reporting. Singh et al. \cite{SDSRG17} propose a novel text analysis based computational approach to automatically detect fake news articles, and they also release a public dataset of valid new articles. Tacchini et al. \cite{TBDMA17} present a technical report on fake news detection with various classification models, and a comprehensive review of detecting spam and rumor is presented by Shu et al. in \cite{SSWTL17}. In this paper, we are the first to provide the systematic formulation of fake news detection problems, illustrate the fake news presentation and factual defects, and introduce unified frameworks for fake news article and creator detection tasks based on deep learning models and heterogeneous network analysis techniques. \\

\noindent \textbf{Spam Detection Research and Applications}: Spams usually denote unsolicited messages or emails with unconfirmed information sent to a large number of recipients on the Internet. The concept \textit{web spam} was first introduced by Convey in \cite{C96} and soon became recognized by the industry as a key challenge \cite{HMS02}. Spam on the Internet can be categorized into \textit{content spam} \cite{DS05, M94, RZT04}, \textit{link spam} \cite{GG05, AM05, ZP07}, \textit{cloaking and redirection} \cite{CC05, WD05, WD06, L09}, and \textit{click spam} \cite{R07, DSYW08, DS07, PZCG09, IJMT05}. Existing detection algorithms for these spams can be roughly divided into three main groups. The first group involves the techniques using content based features, like word/language model \cite{FMN04, NNMF06, SWBR07} and duplicated content analysis \cite{FMN03, FMN05, ULF06}. The second group of techniques mainly rely on the graph connectivity information \cite{CDGMS07, GS07, GWL07, GLZ09}, like link-based trust/distrust propagation \cite{PBMW98, GGP04, KR06}, pruning of connections \cite{BH98, LM01, NOHI04}. The last group of techniques use data like click stream \cite{R07, DSYW08}, user behavior \cite{LGLZMHL08, LZMR08}, and HTTP session information \cite{WCP08} for spam detection. The differences between fake news and conventional spams have been clearly illustrated in Section~\ref{sec:introduction}, which also make these existing spam detection techniques inapplicable to detect fake news articles.\\

\noindent \textbf{Deep Learning Research and Applications}: The essence of deep learning is to compute hierarchical features or representations of the observational data \cite{GBC16, LBH15}. With the surge of deep learning research and applications in recent years, lots of research works have appeared to apply the deep learning methods, like deep belief network \cite{HOT06}, deep Boltzmann machine \cite{SH09}, Deep neural network \cite{J02, KSH12} and Deep autoencoder model \cite{VLLBM10}, in various applications, like speech and audio processing \cite{DHK13, HDYDMJSVNSK12}, language modeling and processing \cite{ASKR12, MH09}, information retrieval \cite{H12, SH09}, objective recognition and computer vision \cite{LBH15}, as well as multimodal and multi-task learning \cite{WBU10, WBU11}.
\section{Conclusion}\label{sec:conclusion}

In this paper, we have studied the fake news article, creator and subject detection problem. Based on the news augmented heterogeneous social network, a set of explicit and latent features can be extracted from the textual information of news articles, creators and subjects respectively. Furthermore, based on the connections among news articles, creators and news subjects, a deep diffusive network model has been proposed for incorporate the network structure information into model learning. In this paper, we also introduce a new diffusive unit model, namely GDU. Model GDU accepts multiple inputs from different sources simultaneously, and can effectively fuse these input for output generation with content ``forget'' and ``adjust'' gates. Extensive experiments done on a real-world fake news dataset, i.e., PolitiFact, have demonstrated the outstanding performance of the proposed model in identifying the fake news articles, creators and subjects in the network.

\vspace{-10pt}
\label{sec:ack}
\section{Acknowledgement}

This work is also supported in part by NSF through grants IIS-1763365 and IIS-1763325.

\balance
\bibliographystyle{plain}
\bibliography{12_reference}

\end{document}